\documentclass[twocolumn,pra,showpacs,superscriptaddress,amssymb,amsmath,amsmath]{revtex4-1}
\usepackage{graphicx}
\usepackage{epstopdf}
\usepackage{comment}
\usepackage{hyperref}
\usepackage{bm}

\hypersetup{pdfstartpage=1,pdfmenubar=true,pdfpagemode=None,pdftoolbar=true,
urlcolor=blue,linkcolor=blue,colorlinks = true,citecolor=blue, bookmarksopen=false}

\begin{document}
\title{Interactions and charge transfer dynamics of an Al$^+$ ion immersed \\in ultracold Rb and Sr atoms}

\author{Micha\l~Tomza}
\email{michal.tomza@fuw.edu.pl}
\affiliation{Faculty of Physics, University of Warsaw, Pasteura 5, 02-093 Warsaw, Poland}
\author{Mateusz Lisaj}
\affiliation{Faculty of Physics, University of Warsaw, Pasteura 5, 02-093 Warsaw, Poland}
 
\date{\today}

\begin{abstract}

Atomic clocks based on an Al$^+$ ion sympathetically cooled by a laser-cooled alkaline-earth ion have achieved unprecedented accuracy. Here, we investigate theoretically interactions and charge transfer dynamics of an Al$^+$ ion immersed in an ultracold gas of Rb and Sr atoms. We calculate potential energy curves and transition electric dipole moments for the (Al+Rb)$^+$ and (Al+Sr)$^+$ ion-atom systems using coupled cluster and multireference configuration interaction methods with scalar relativistic effects included within the small-core energy-consistent pseudopotentials in Rb and Sr atoms. The long-range interaction coefficients are also reported. We use the electronic structure data to investigate cold collisions and charge transfer dynamics. Scattering of an Al$^+$ ion with alkali-metal or alkaline-earth-metal atom is governed by one potential energy curve whereas charge transfer can lead to several electronic states mixed by the relativistic spin-orbit coupling. We examine the branching ratios resulting from the interplay of the short- and long-range effects, as well as the prospects for the laser-field control and formation of molecular ions. We propose to employ the atomic clock transition in an Al$^+$ ion to monitor ion-atom scattering dynamics \textit{via} quantum logic spectroscopy. The presented results pave the way for the application of atomic ions other than alkali-metal and alkaline-earth-metal ones in the field of cold hybrid ion-atom experiments.

\end{abstract}

\maketitle

\section{Introduction}

Trapped and laser-cooled atomic ions allow for tackling questions touching upon the very fundamentals of quantum mechanics~\cite{WinelandRMP13}. They have found many applications ranging from quantum simulation and quantum computation~\cite{HaeffnerPR08,BlattNatPhys12} to quantum metrology and sensing~\cite{Wineland02,DegenRMP17,KozlovRMP18}. For example, optical atomic clocks based on single trapped ions have been realized~\cite{SchmidtScience05,RosenbandScience08,ChouPRL10,HuntemannPRL16}. Recently, the most stable and accurate single-ion atomic clock with fractional uncertainty as small as $10^{-18}$ has been constructed using the $^1S_0$-$^3P_0$ transition in an Al$^+$ ion sympathetically cooled by a co-trapped and laser-cooled Mg$^+$ ion~\cite{ChenPRL17,Brewer2019a}.

In the last decade, a new field of cold hybrid systems has emerged, where laser-cooled trapped ions are combined in a single experimental setup with ultracold atoms~\cite{HarterCP14,CoteAAMOP16,TomzaRMP18}. Ultracold atoms on its own have been used in numerous ground-breaking experiments allowing to explore a plethora of quantum phenomena~\cite{BlochRMP08,GrossScience17}. Mixtures of laser-cooled trapped ions and ultracold atoms may combine the best features of two well established fields of research and promise new exicting applications in quantum physics and chemistry~\cite{TomzaRMP18}. They can be employed to study cold collisions and sympathetic cooling~\cite{CotePRA00,ZipkesNature10,RaviNatCommun12,HoltkemeierPRL16,PetrovJCP17,Feldker2019}, chemical reactions~\cite{HallPRL12}, spin and charge-transfer dynamics of ionic impurities~\cite{MakarovPRA03,RatschbacherNatPhys12,FurstPRA18,SikorskyPRL18,CotePRL18,SaitoPRA17}, spin-controlled ion-atom chemistry~\cite{SikorskyNC18}, simulation of solid-state physics~\cite{BissbortPRL13}, and quantum computation~\cite{DoerkPRA10}. Until now, most of cold experiments have used alkaline-earth-metal ions trapped and laser-cooled in Paul traps immersed into neutral alkali-metal or alkaline-earth-metal atoms trapped in magneto-optical, magnetic, or dipole traps. Investigated cold atomic combinations include: Yb$^+$/Yb~\cite{GrierPRL09}, Ca$^+$/Rb~\cite{HallMP13a}, Ba$^+$/Ca~\cite{SullivanPRL12}, Yb$^+$/Ca~\cite{RellergertPRL11}, Yb$^+$/Rb~\cite{ZipkesNature10}, Ca$^+$/Li~\cite{HazePRA13}, Ca$^+$/Rb~\cite{HallPRL11}, Ca$^+$/Na~\cite{SmithAPB14}, Yb$^+$/Li~\cite{JogerPRA17}, Sr$^+$/Rb~\cite{MeirPRL16}, Rb$^+$/Rb~\cite{RaviNatCommun12,HartePRL12}. Additionally, sympathetically cooled molecular ions such as N$_2^+$~\cite{HallPRL12} and OH$^-$~\cite{DeiglmayrPRA12} were immersed into ultracold Rb atoms. Recently, the direct formation of ions inside ultacold gases via the ionization of Rydberg states was also achieved~\cite{JyothiPRL16,KleinbachPRL18,SchmidPRL18}.

\begin{figure}[b]
\begin{center}
\includegraphics[width=0.7\columnwidth]{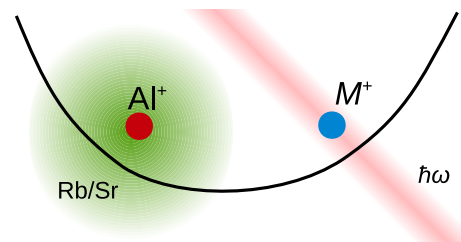}
\end{center}
\caption{Schematic representation of the considered experimental arrangement with an Al$^+$ ion co-trapped in a Paul trap with a laser-cooled alkaline-earth ion $M^+$. An Al$^+$ ion is overlapped with a small cloud of ulracold Rb or Sr atoms, whereas an alkaline-earth ion is optically addressed to control and read a quantum state of both ions.}
\label{fig:scheme}
\end{figure}

In this paper, we propose to employ an Al$^+$ ion co-trapped in a Paul trap with a laser-cooled alkaline-earth ion and overlapped with a small cloud of ulracold Rb or Sr atoms to investigate cold interactions and collisions, including charge-transfer dynamics, between an Al$^+$ ion and Rb or Sr atoms (see~Fig.~\ref{fig:scheme}). Rb and Sr are selected as a prototype alkali-metal and alkaline-earth-metal atoms. Quantum state detection and preparation of the Al$^+$ ion may be realized by optical addressing of a co-trapped alkaline-earth ion within the quantum logic spectroscopy~\cite{SchmidtScience05}. Prospects for molecular ions formation are also analyzed. Scattering calculations are based on the electronic structure data, including potential energy curves and transition electric dipole moments, which we calculate using \textit{ab initio} electronic structure approaches such as the coupled cluster and multireference configuration interaction methods.

The plan of this paper is as follows. Section~\ref{sec:theory} describes the used theoretical electronic structure and quantum scattering methods. Section~\ref{sec:results} presents and analyzes the electronic and collisional properties of the investigated ion-atom systems. Section~\ref{sec:summary} summarizes our paper and discusses future possible applications.

\section{Computational details}
\label{sec:theory}

An Al$^+$ ion has a closed-shell electronic ground state of the $^1S_0$ symmetry. Its interaction with a ground-state alkali-metal (alkaline-earth-metal) atom in the $^2S_{1/2}$ ($^1S_{0}$) electronic state results in a single molecular electronic state of the $^2\Sigma^+$ ($^1\Sigma^+$) symmetry. The ionization potential of an Al atom is larger than of all alkali-metal atoms and of Sr and Ba among alkaline-earth-metal atoms, therefore the charge transfer process is energetically possible in collisions between an Al$^+$ ion and those atoms~\cite{TomzaPRA15b}. Interestingly, such a process leads to several electronic states dissociating to the lowest ion-atom asymptotes with an Al atom in the $^2P_{1/2}$ or $^2P_{3/2}$ electronic state. In the non-relativistic picture, if an Al atom interacts with an alkali-metal ion then two molecular electronic states of the $^2\Sigma^+$ and $^2\Pi$ symmetries exist, and if an Al atom interacts with an alkaline-earth-metal ion then four molecular electronic states of the $^1\Sigma^+$, $^1\Pi$, $^3\Sigma^+$, and $^3\Pi$ symmetries are possible. Relativistic effects, namely the spin-orbit coupling which splits the atomic $^2P$ electronic state into the $^2P_{1/2}$ and $^2P_{3/2}$ fine states, couple and mix different molecular electronic states. Relativistic effects, which are of moderate values in the considered systems, can be included perturbatively~\cite{TomzaPCCP11,TomzaPRA12}.
 
To calculate potential energy curves in the Born-Oppenheimer approximation we adopt the computational scheme successfully applied to the ground and excited electronic states of the (LiYb)$^+$ molecular ion~\cite{TomzaPRA15a} and SrYb molecule~\cite{TomzaPCCP11}. The non-relativistic potential energy curves are obtained using the multireference configuration interaction method restricted to single and double exictations with the Davidson correction, MRCISD+Q, starting from orbitals obtained with the multi-configurational self-consistent field method, MCSCF~\cite{WernerJCP88}. Complete active spaces constructed from $3s3p$, $5s5p$, and $5s5p4d$ atomic orbitals of Al, Rb, and Sr are used.
 When possible the spin-restricted open-shell coupled cluster method restricted to single, double, and noniterative triple excitations, starting from the restricted open-shell Hartree-Fock (ROHF) orbitals, RCCSD(T), is employed~\cite{KnowlesJCP93}. The interaction energies are obtained with the supermolecular method with the basis set superposition error corrected by using the counterpoise correction~\cite{BoysMP70}
\begin{equation}
V_{(\textrm{Al}+X)^+}=E_{(\text{Al}+X)^+}-E_{\text{Al}^+}-E_X\,,
\end{equation}
where $E_{(\textrm{Al}+X)^+}$ is the total energy of the ion-atom system, and $E_{\text{Al}^+}$ and $E_X$ are the total energies of the Al$^+$ ion and $X$ atom computed in the diatom basis set.

The Al atom is described by the augmented correlation consistent polarized valence and core-valence quintuple-$\zeta$ quality basis sets, aug-cc-pV5Z (in MRCISD+Q calculations) and aug-cc-pCV5Z (in RCCSD(T) calculations), respectively~\cite{WoonJCP93}. The scalar relativistic effects in Rb and Sr atoms are included by employing the small-core relativistic energy-consistent pseudopotentials (ECP) to replace the inner-shells electrons~\cite{DolgCR12}. The pseudopotentials from the Stuttgart library are employed in all calculations. The Rb and Sr atoms are  described with the ECP28MDF pseudopotentials~\cite{LimJCP05,LimJCP06} together with the $[14s14p7d6f1g]$ and $[14s11p6d5f4g]$ basis sets reported in Refs.~\cite{TomzaPCCP11,TomzaMP13}. The atomic basis sets are additionally augmented in all calculations by the set of $[3s3p2d1f1g]$ bond functions to accelerate the convergence towards the complete basis set limit~\cite{midbond}. All electronic structure calculations are performed with the \textsc{Molpro} package of \textit{ab initio} programs~\cite{Molpro}.

The relativistic potential energy curves $V_{|\Omega|}(R)$ associated with the two lowest atomic asymptotes of the Al($^2P_J$) atom interacting with the Rb$^+$($^1S_0$) or Sr$^+$($^2S_{1/2}$) ion are obtained within the perturbation theory by diagonalizing the interaction Hamiltonians with the spin-orbit coupling included between non-relativistic curves. Relativistic states are characterized by the projection of the total angular momentum on the molecular axis $|\Omega|$.

For Al($^2P_J$)+Rb$^+$($^1S_0$), the relativistic molecular potential energy curves are given by
\begin{equation}\label{eq:12_RbAl+}
V_{|\Omega|=\frac{1}{2}}(R)=\begin{pmatrix} V_{^2\Sigma}(R) & A^{^2\Sigma/^2\Pi}_\text{SO}(R)\\  A^{^2\Sigma/^2\Pi}_\text{SO}(R) & V_{^2\Pi}(R) - A^{^2\Pi/^2\Pi}_\text{SO}(R) \end{pmatrix}\,,
\end{equation}
\begin{equation}\label{eq:32_RbAl+}
V_{|\Omega|=\frac{3}{2}}(R)=V_{^2\Pi}(R)+ A^{^2\Pi/^2\Pi}_\text{SO}(R)\,,
\end{equation}
where $V_{^{2S+1}|\Lambda|}(R)$ is the non-relativistic potential energy curve for the $^{2S+1}|\Lambda|$ electronic state and $A^{^{2S+1}|\Lambda|/^{2S'+1}|\Lambda'|}_\text{SO}(R)$ is the matrix element of the spin-orbit coupling Hamiltonian, $\hat{H}_\text{SO}=\hat{S}\cdot\hat{L}$, between the $^{2S+1}|\Lambda|$ and $^{2S'+1}|\Lambda'|$ electronic states~\cite{Dyall2007}. 

For Al($^2P_J$)+Sr$^+$($^2S_{1/2}$), the relativistic molecular potential energy curves are given by
\begin{equation}
V_{|\Omega|=0^+}(R)=\begin{pmatrix} V_{^1\Sigma}(R) & A^{^1\Sigma/^3\Pi}_\text{SO}(R)\\  A^{^1\Sigma/^3\Pi}_\text{SO}(R) & V_{^3\Pi}(R) - A^{^3\Pi/^3\Pi}_\text{SO}(R) \end{pmatrix}\,,
\end{equation}
\begin{equation}
V_{|\Omega|=1}(R)=\begin{pmatrix} V_{^3\Sigma}(R) & A^{^3\Sigma/^3\Pi}_\text{SO}(R) & A^{^3\Sigma/^1\Pi}_\text{SO}(R) \\
A^{^3\Sigma/^3\Pi}_\text{SO}(R) & V_{^3\Pi}(R) & -A^{^3\Pi/^1\Pi}_\text{SO}(R) \\
A^{^3\Sigma/^1\Pi}_\text{SO}(R) & -A^{^3\Pi/^1\Pi}_\text{SO}(R) & V_{^1\Pi}(R)  \end{pmatrix}\,,
\end{equation}
\begin{equation}
V_{|\Omega|=2}(R)=V_{^3\Pi}(R)+A^{^3\Pi/^3\Pi}_\text{SO}(R)\,.
\end{equation}

Asymptotically, the spin-orbit coupling matrix elements are related to the fine splitting between $^2P_{1/2}$ and $^2P_{3/2}$ electronic states of the Al atom $\Delta E_\text{fs}^{\text{Al}}=112.1\,$cm$^{-1}$~\cite{nist}. $A^{^2\Pi/^2\Pi}_\text{SO}(R\to\infty)=A^{^3\Sigma/^3\Pi}_\text{SO}(R\to\infty)=A^{^3\Sigma/^1\Pi}_\text{SO}(R\to\infty)=A^{^3\Pi/^1\Pi}_\text{SO}(R\to\infty)=\Delta E_\text{fs}^{\text{Al}}/3$ and $A^{^2\Sigma/^2\Pi}_\text{SO}(R\to\infty)=A^{^1\Sigma/^3\Pi}_\text{SO}(R\to\infty)=\sqrt{2}\Delta E_\text{fs}^{\text{Al}}/3$. In the present study, we neglect the $R$-dependence of the matrix elements of the spin-orbit coupling assuming their asymptotic values. This approximation should not significantly affect the presented results dominated by the dynamics at intermediate- and long-range distances.

The interaction potential between the ion and the atom at large internuclear distances $R$ is generally given by
\begin{equation}\label{eq:long-range}
V_{(\textrm{Al}+X)^+}(R)=-\frac{C^\textrm{elst}_3}{R^3}-\frac{C^\textrm{ind}_4}{R^4}-\frac{C^\textrm{ind}_6}{R^6}-\frac{C^\textrm{disp}_6}{R^6}+\dots\,,
\end{equation}  
where $-{C^\textrm{elst}_3}/{R^3}$ term describes the electrostatic interaction between the charge of the ion and the permanent quadrupole moment of the atom with non-zero electronic angular momentum ($L\neq 0$), $-{C^\textrm{ind}_4}/{R^4}-{C^\textrm{ind}_6}/{R^6}$ terms describe the induction interaction between the charge of the ion and the induced electric dipole and quadruple moments of the atom, respectively, and $-{C^\textrm{disp}_6}/{R^6}$ term  describes the dispersion interaction between instantaneous dipole-induced dipole moments of the ion and atom arising
due to quantum fluctuations~\cite{JeziorskiCR94}. 

The leading long-range interaction coefficients do not depend on the total electronic spin of molecular electronic states and for $\Sigma$ and $\Pi$ symmetries they are given by
\begin{equation}
\begin{split}
C^\textrm{elst}_3(\Sigma)&= q\Theta_\text{at}\,,\\
C^\textrm{elst}_3(\Pi)&= -\frac{1}{2}q\Theta_\text{at}\,,\\
C^{\mathrm{ind}}_4(\Sigma)&=\frac{1}{2}q^2\big(\bar{\alpha}_\textrm{at}+\frac{2}{3}\Delta\alpha_\textrm{at}\big)\,,\\
C^{\mathrm{ind}}_4(\Pi)&=\frac{1}{2}q^2\big(\bar{\alpha}_\textrm{at}-\frac{1}{3}\Delta\alpha_\textrm{at}\big)\,,\\
C^{\mathrm{ind}}_6(\Sigma)&=\frac{1}{2}q^2\big(\bar{\beta}_\textrm{at}+\frac{2}{3}\Delta\beta_\textrm{at}\big)\,,\\
C^{\mathrm{ind}}_6(\Pi)&=\frac{1}{2}q^2\big(\bar{\beta}_\textrm{at}-\frac{1}{3}\Delta\beta_\textrm{at}\big)\,,\\
C^{\mathrm{disp}}_6(\Sigma)&=\frac{3}{\pi}\int^\infty_0\alpha_\text{ion}(i\omega)\big(\bar{\alpha}_\textrm{at}(i\omega)+\frac{2}{3}\Delta\alpha_\textrm{at}(i\omega)\big)d\omega\,,\\
C^{\mathrm{disp}}_6(\Pi)&=\frac{3}{\pi}\int^\infty_0\alpha_\text{ion}(i\omega)\big(\bar{\alpha}_\textrm{at}(i\omega)-\frac{1}{3}\Delta\alpha_\textrm{at}(i\omega)\big)d\omega\,,\\
\end{split}
\end{equation}
where $q$ is the charge of the ion, $\Theta_\text{at}$ is the permanent quadrupole moment of the atom, $\bar{\alpha}_\textrm{at}=(\alpha_{xx}+\alpha_{yy}+\alpha_{zz})/3$ and $\Delta\alpha_\text{at}=\alpha_{zz}-\alpha_{xx}$ are the isotropic (scalar) and anisotropic (tensor) components of the static electric dipole polarizability of the atom, $\bar{\beta}_\textrm{at}$ and $\Delta{\beta}_\textrm{at}$ are the components of the static electric quadrupole polarizability of the atom, and $\bar{\alpha}_\text{ion(atom)}(i\omega)$ and $\Delta\alpha_\text{ion(atom)}(i\omega)$ are the components of the dynamic polarizbility of the ion(atom) at imaginary frequency. The permanent quadrupole moment and static quadrupole polarizability of the Al atom are calculated with the RCCSD(T) and finite field methods. The dynamic electric dipole polarizabilities at imaginary frequency of the Rb and Sr atoms are taken from Ref.~\cite{DerevienkoADNDT10}, whereas the dynamic polarizability of the Al$^+$ and Rb$^+$ ions are obtained by using the explicitly connected representation of the expectation value and polarization propagator within the coupled cluster method~\cite{KoronaMP06}. The dynamic polarizabilities of the Sr$^+$ ion and the Al atom are obtained from the sum over state expression using the transition moments and energy levels from the NIST Atomic Spectra Database~\cite{nist}. 

Rate constants for elastic scattering and inelastic charge-exchange reactive collisions are calculated as implemented and described in Refs.~\cite{TomzaPRA15a,TomzaPRA15b,TomzaPRL14}. The time-independent Schr\"odinger equation for the nuclear motion of colliding ion and atom is solve using the renormalized Numerov propagator~\cite{JohnsonJCP78} with step-size doubling. The wave functions are propagated to  large interatomic separations and the $K$ and $S$ matrices are extracted by imposing the long-range scattering boundary conditions in terms of the Bessel functions. The elastic rate constants and scattering lengths are obtained from the $S$ matrix for the entrance channel, while inelastic rate constants are calculated using Fermi golden rule type expressions based on Einstein coefficients between bound and continuum ro-vibrational wave functions of relevant electronic states.
{We neglect the hyperfine structure of Rb, Al, and Sr$^+$ because it should be negligible for the scattering dynamics in the entrance channels, while studying interplay of the fine and hyperfine structures in the exit channels is out of the scope of this paper.}

\section{Numerical results and discussion}
\label{sec:results}

\subsection{Potential energy curves and transition electric dipole moments}

\begin{figure}[tb]
\begin{center}
\includegraphics[width=\columnwidth]{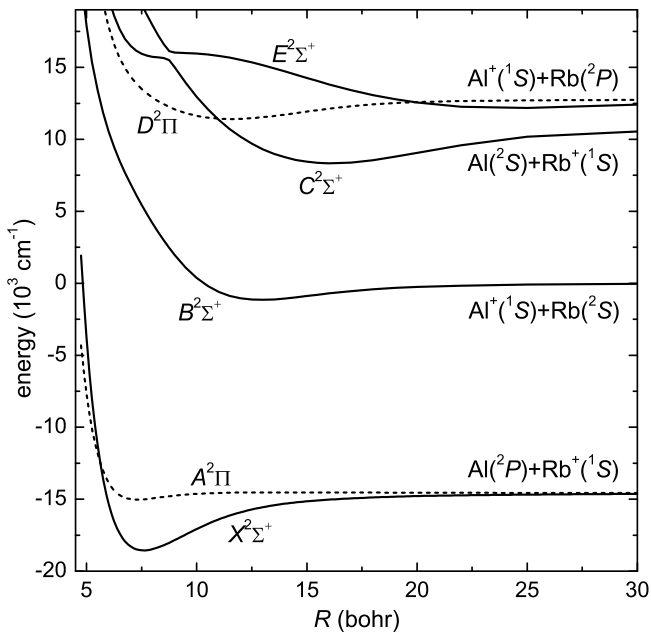}
\end{center}
\caption{Non-relativistic molecular potentials energy curves of the (Al+Rb)$^+$ ion-atom system.}
\label{fig:V_RbAl+}
\end{figure}

\begin{figure}[tb!]
\begin{center}
\includegraphics[width=\columnwidth]{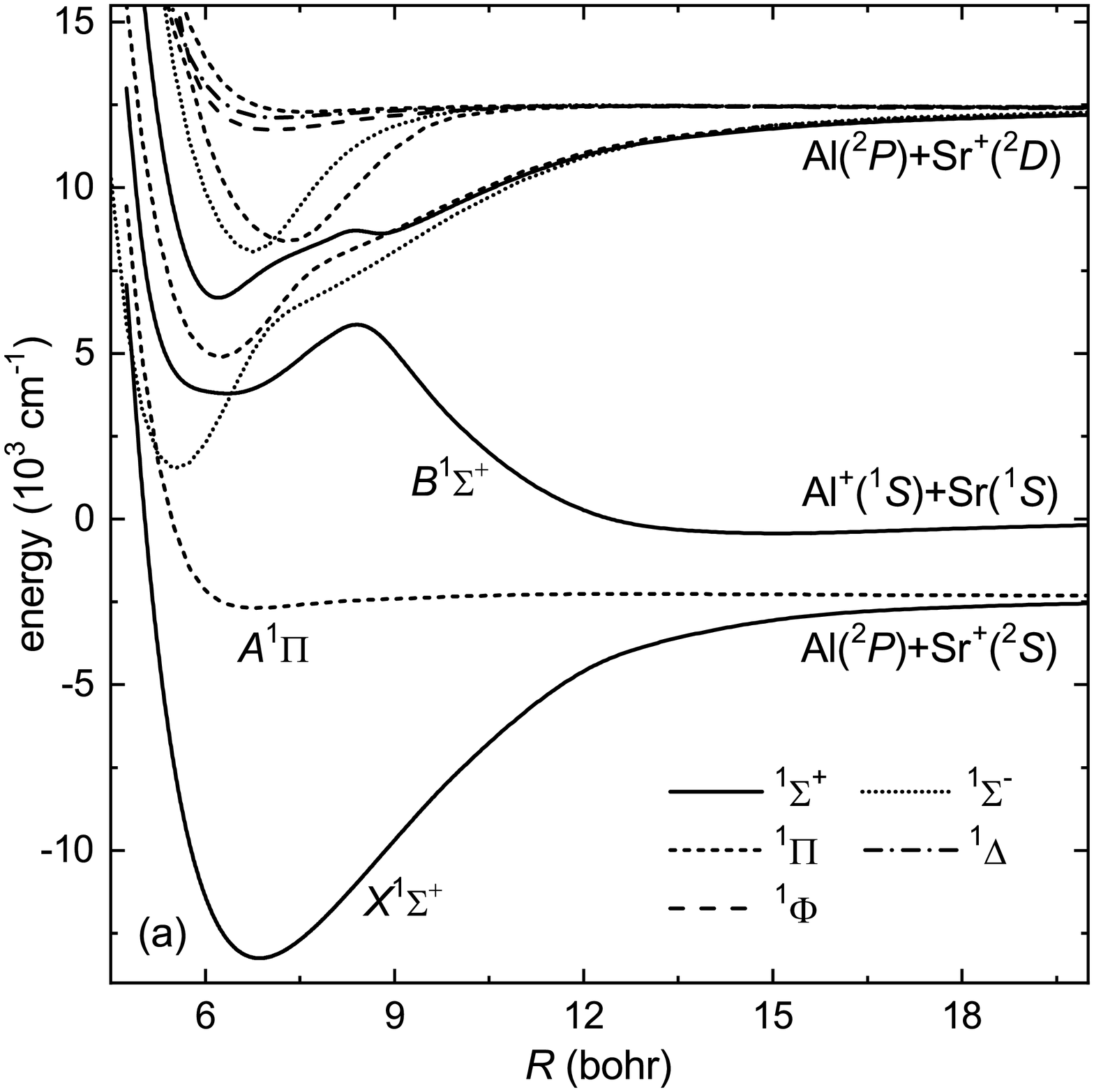}
\includegraphics[width=\columnwidth]{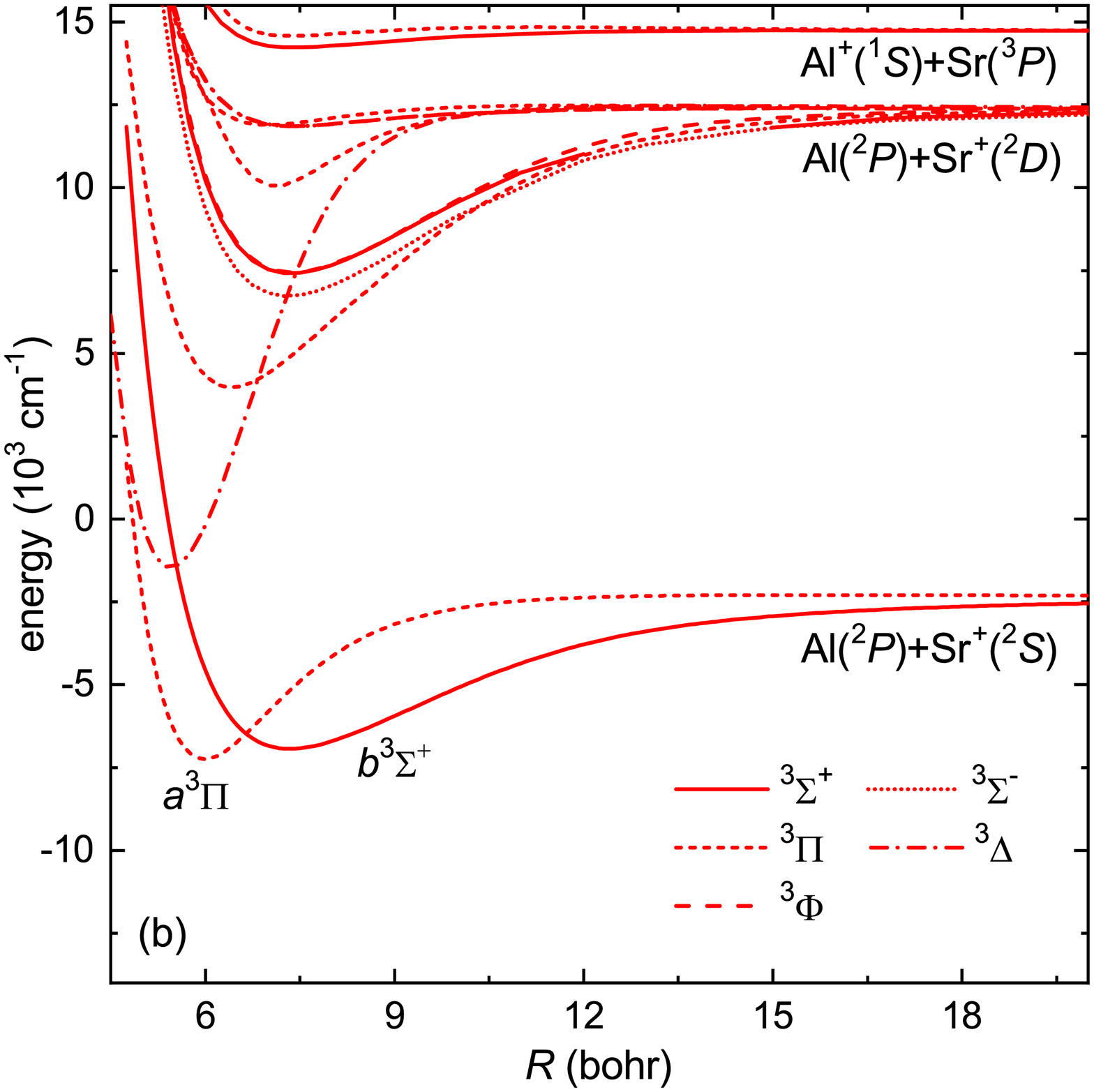}
\end{center}
\caption{Non-relativistic singlet (a) and triplet (b) molecular potentials energy curves of the (Al+Sr)$^+$ ion-atom system.}
\label{fig:V_SrAl+}
\end{figure}

\begin{figure}[tb!]
\begin{center}
\includegraphics[width=0.97\columnwidth]{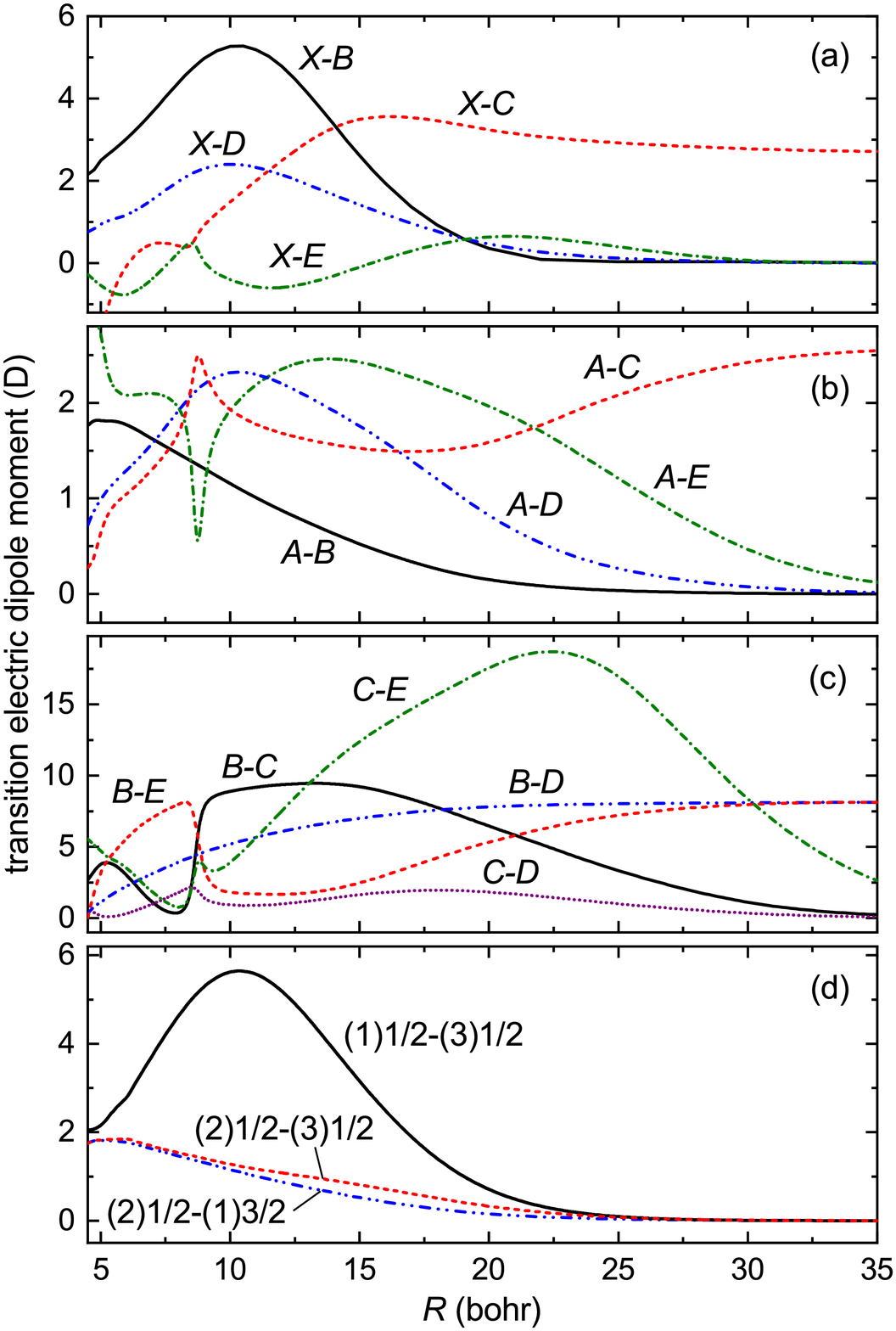}
\end{center}
\caption{Transition electric dipole moments between the ground and excited molecular electronic states of the (Al+Rb)$^+$ ion-atom system. Panels (a)-(c) are for transitions between non-relativistic states and panel (d) is for transitions between relativistic states.}
\label{fig:dm_RbAl+}
\end{figure}

\begin{figure}[tb!]
\begin{center}
\includegraphics[width=0.97\columnwidth]{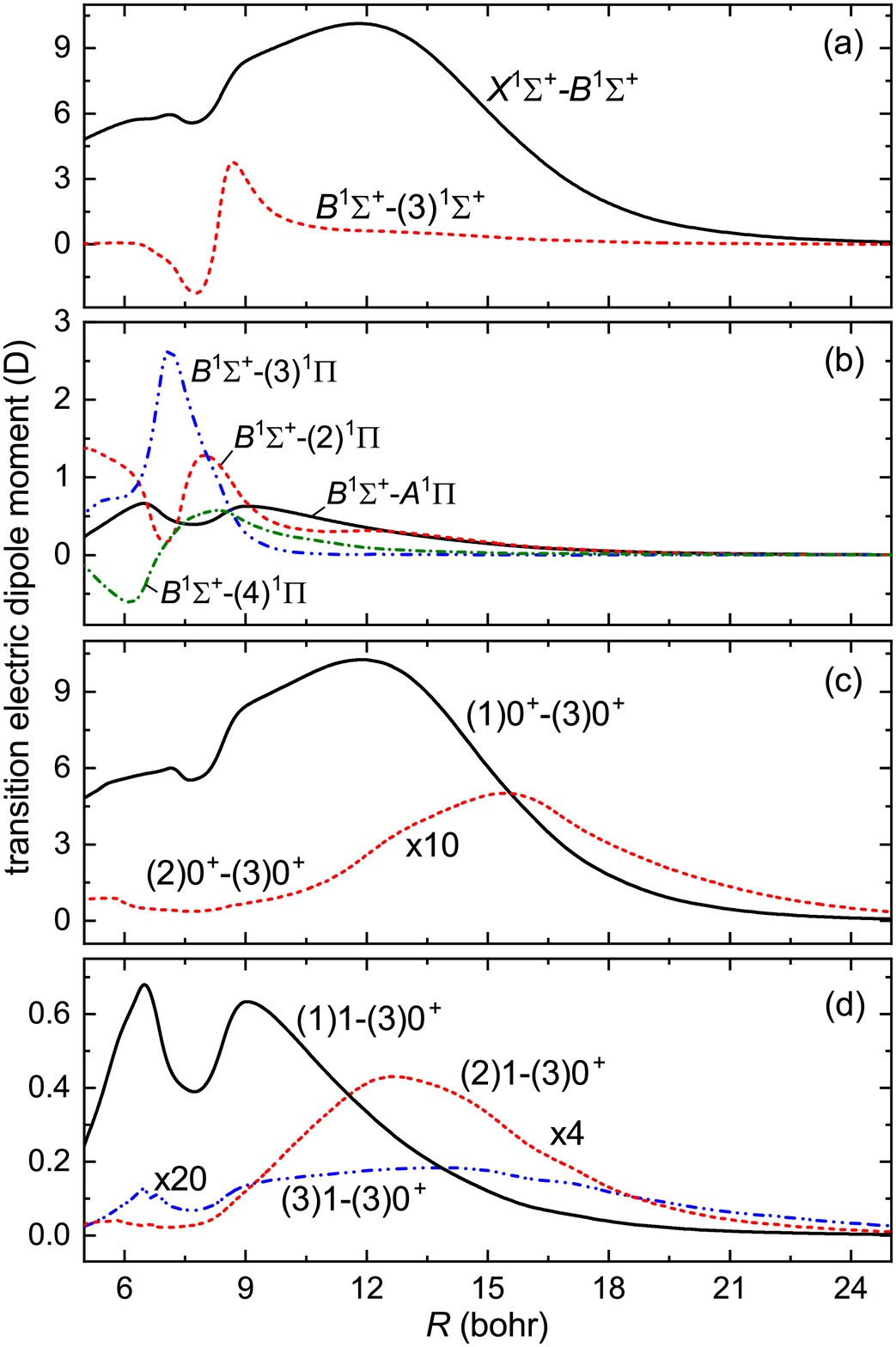}
\end{center}
\caption{Transition electric dipole moments between selected ground and excited molecular electronic states of the (Al+Sr)$^+$ ion-atom system.  Panels (a)-(b) are for transitions between non-relativistic states and panels (c)-(d) are for transitions between relativistic states.
}
\label{fig:dm_SrAl+}
\end{figure}

Figures~\ref{fig:V_RbAl+} and~\ref{fig:V_SrAl+} present non-relatvistic molecular potential energy curves for the ground and excited electronic states of the (Al+Rb)$^+$ and (Al+Sr)$^+$ ion-atom systems, respectively. Their spectroscopic characteristics, such as equilibrium bond lengths $R_e$, well depths $D_e$, harmonic constants $\omega_e$, and rotational constants $B_e$, are collected in Tables~\ref{tab:spec_AlRb+} and~\ref{tab:spec_AlSr+}. Figures~\ref{fig:dm_RbAl+} and~\ref{fig:dm_SrAl+} show the transition electric dipole moments between the ground and excited molecular electronic states of the (Al+Rb)$^+$ and (Al+Sr)$^+$ ion-atom systems, respectively. The calculated electronic structure data are available in numerical form from the authors upon request.

All relevant non-relativistic molecular electronic states of the (Al+Rb)$^+$ system, presented in Fig.~\ref{fig:V_RbAl+}, are of the $^2\Sigma^+$ or $^2\Pi$ symmetry. The entrance $B^2\Sigma^+$ electronic state, related to the Al$^+$($^1S$)+Rb($^2S$) atomic threshold, is well separated from lower and higher lying electronic states. This suggests that non-radiative charge transfer, which could be driven by the non-adiabatic radial couplings~\cite{RellergertPRL11,SayfutyarovaPRA13,LiPRA19}, will be suppressed and negligible as compared to the radiative charge transfer, which is driven by the transition electric dipole moments. The spontaneous radiative transitions to the electronic states related to the Al($^2P$)+Rb$^+$($^1S$) atomic threshold can result in the radiative charge transfer or the radiative association depending on the structure of potential energy curves and transition electric dipole moments. In the case of the (Al+Rb)$^+$ system, as shown in Fig.~\ref{fig:dm_RbAl+}, the transition moment between the $B^2\Sigma^+$ and $X^2\Sigma^+$ electronic states is noticeably larger than the one between the $B^2\Sigma^+$ and $A^2\Pi$ states, in favor of radiative processes leading to the $X^2\Sigma^+$ state. The equilibrium distance of the $B^2\Sigma^+$ state is larger as compared to the $X^2\Sigma^+$ and $A^2\Pi$ states but some overlap between vibrational states associated with these electronic states can be expected.

\begin{table}[tb]
\caption{Spectroscopic characteristics of molecular electronic states of the ($^{27}$Al+$^{85}$Rb)$^+$ ion-atom system: equilibrium bond lengths $R_e$, well depths $D_e$, harmonic constants $\omega_e$, and rotational constants $B_e$. Results obtained with the couple-cluster (CC) and configuration interaction (CI) methods are presented as described in the text. \label{tab:spec_AlRb+}} 
\begin{ruledtabular}
\begin{tabular}{llrrrr}
State & Met. & $R_e\,$(bohr) & $D_e\,$(cm$^{-1}$) &  $\omega_e\,$(cm$^{-1}$) & $B_e\,$(cm$^{-1}$) \\
\hline
\multicolumn{6}{c}{Al(${}^2P$)+Rb$^+$(${}^1S$)} \\
$X^2\Sigma^+$ & CC & 7.59 & 3971 & 85.3 & 0.0510 \\
$X^2\Sigma^+$ & CI & 7.66 & 3940 & 83.3 & 0.0502 \\
$A^2\Pi$ & CC & 7.31 & 443 & 56.9 & 0.0550 \\
$A^2\Pi$ & CI & 7.35 & 395 & 55.5 & 0.0544 \\
\hline
\multicolumn{6}{c}{Al$^+$(${}^1S$)+Rb(${}^2S$)} \\
$B^2\Sigma^+$ & CI & 12.91 & 1158 & 34.6 & 0.0176 \\
\hline
\multicolumn{6}{c}{Al(${}^2S$)+Rb$^+$(${}^1S$)} \\
$C^2\Sigma^+$ & CI & 16.13 & 2425 & 30.0 & 0.0113  \\
\hline
\multicolumn{6}{c}{Al$^+$(${}^1S$)+Rb(${}^2P$)} \\
$D^2\Pi$ & CI & 11.56 & 1327 & 29.6 & 0.0220 \\
$E^2\Sigma^+$ & CI & 24.24 &  559 & 13.2 & 0.00501 \\
\end{tabular}
\end{ruledtabular}
\end{table}

\begin{table}[tb!]
\caption{Spectroscopic characteristics of molecular electronic states of the ($^{27}$Al+$^{88}$Sr)$^+$ ion-atom system: equilibrium bond lengths $R_e$, well depths $D_e$, harmonic constants $\omega_e$, and rotational constants $B_e$. Results obtained with the couple-cluster (CC) and configuration interaction (CI) methods are presented as described in the text. \label{tab:spec_AlSr+}} 
\begin{ruledtabular}
\begin{tabular}{llrrrr}
State & Met. & $R_e\,$(bohr) & $D_e\,$(cm$^{-1}$) &  $\omega_e\,$(cm$^{-1}$) & $B_e\,$(cm$^{-1}$) \\
\hline
\multicolumn{6}{c}{Al(${}^2P$)+Sr$^+$(${}^2S$)} \\
$X^1\Sigma^+$ & CC & 6.84 & 10949 & 142.2 & 0.0628  \\
$X^1\Sigma^+$ & CI & 7.07 &  9763 & 131.5 & 0.0582   \\
$a^3\Pi$ & CC & 6.00 & 4897 & 180.2 & 0.0818 \\
$a^3\Pi$ & CI & 6.26 & 3951 & 165.9 & 0.0745 \\
$b^3\Sigma^+$ & CC & 7.35 &  4586 & 89.0 & 0.0545 \\
$b^3\Sigma^+$ & CI & 7.51 &  3964 & 79.7 & 0.0521\\
$A^1\Pi$ & CI & 6.77 & 341 & 67.6 & 0.0602 \\
\hline
\multicolumn{6}{c}{Al$^+$(${}^1S$)+Sr(${}^1S$)} \\
$B^1\Sigma^+$ & CI & 14.55 & 499 & 26.0 & 0.0140   \\
\hline
\multicolumn{6}{c}{Al(${}^2P$)+Sr$^+$(${}^2D$)} \\
$(3)^1\Sigma^+$ & CI & 6.17 & 5796 & 213.1 & 0.0765\\
$(1)^1\Sigma^-$ & CI & 5.54 & 10894 & 233.8 & 0.0951  \\
$(2)^1\Sigma^-$ & CI & 6.75 &  4324 & 195.4 & 0.0640 \\
$(2)^1\Pi$ & CI & 6.22 & 7521 & 176.5 & 0.0753 \\
$(3)^1\Pi$ & CI & 7.30 & 4015 & 134.1 & 0.0547 \\
$(4)^1\Pi$ & CI & 7.57 & 106 & 44.4 &  0.0509 \\
$(1)^1\Delta$   & CI & 7.11 & 287 & 60.5 & 0.0576  \\
$(2)^1\Delta$ & CI &  \multicolumn{4}{c}{repulsive} \\
$(1)^1\Phi$ & CI & 7.02 &   660 & 69.9  & 0.0591\\
$(2)^3\Sigma^+$ & CI & 7.36 & 4985 & 95.3 & 0.0539 \\
$(1)^3\Sigma^-$ & CI & 7.31 & 5664 & 99.1 & 0.0546  \\
$(2)^3\Sigma^-$ & CI & 7.40 & 548 & 56.4 & 0.0533 \\
$(2)^3\Pi$ & CI & 6.40 & 8456 & 141.0 & 0.0713 \\
$(3)^3\Pi$ & CI & 7.11 & 2345 & 141.7 & 0.0578 \\
$(4)^3\Pi$ & CI & 7.00 & 511 & 74.8 & 0.0595 \\
$(1)^3\Delta$ & CI & 5.44 & 13969 & 256.4 & 0.0983 \\
$(2)^3\Delta$ & CI & \multicolumn{4}{c}{repulsive}  \\
$(1)^3\Phi$ & CI & 7.39 & 4951 & 95.3 & 0.0537\\
\hline
\multicolumn{6}{c}{Al$^+$(${}^1S$)+Sr(${}^3P$)} \\
$(4)^3\Sigma^+$ & CI & 7.38 & 474 & 54.0 & 0.0535 \\
$(5)^3\Pi$ & CI & 7.34 & 128 & 54.8 & 0.0541 \\
\end{tabular}
\end{ruledtabular}
\end{table}

\begin{table}[b]
\caption{Electrostatic, induction, and dispersion interaction coefficients (in atomic units) describing the long-range part of the interaction potentials between the Al$^+$ ion and the Rb or Sr atom and between the Rb$^+$ or Sr$^+$ ion and the Al atom, all in the ground electronic state.\label{tab:Cn}} 
\begin{ruledtabular}
\begin{tabular}{lcrrrr}
System & Sym. & $C^{\mathrm{el}}_3$& $C^{\mathrm{ind}}_4$ & $C^{\mathrm{ind}}_6$ &   $C^{\mathrm{disp}}_6$  \\
\hline
Al$^+$(${}^1S$)+Rb(${}^2S$) & $^2\Sigma$ & - & 159.8 & 3346 & 615 \\
Al(${}^2P$)+Rb$^+$(${}^1S$) & $^2\Sigma$ & 5.20 & 37.3 & 414 & 106 \\
Al(${}^2P$)+Rb$^+$(${}^1S$) & $^2\Pi$ & -2.60 & 24.6 & 260 & 70 \\
\hline
Al$^+$(${}^1S$)+Sr(${}^1S$) & $^1\Sigma$ & - & 99.6 & 2311 & 567 \\
Al(${}^2P$)+Sr$^+$(${}^2S$) & $^{1}\Sigma$, $^{3}\Sigma$ & 5.20 & 37.3 & 414 & 578 \\
Al(${}^2P$)+Sr$^+$(${}^2S$) & $^1\Pi$, $^3\Pi$ & -2.60 & 24.6 & 260 & 381 \\
\end{tabular}
\end{ruledtabular}
\end{table}

The electronic structure of the (Al+Sr)$^+$ system, presented in Fig.~\ref{fig:V_SrAl+}, is noticeably more complex as compared to the (Al+Rb)$^+$ system due to one electron more in the valence shell of the Sr atom. Thus, two families of singlet and triplet molecular electronic states of the $^1\Sigma^+$, $^1\Sigma^-$, $^1\Pi$, $^1\Delta$, $^1\Phi$ and $^3\Sigma^+$, $^3\Sigma^-$, $^3\Pi$, $^3\Delta$, $^3\Phi$ symmetries are relevant in the present study. The Al($^2P$)+Sr$^+$($^2D$) atomic threshold is related with the richest family of 18 different non-relativistic molecular electronic states. The inclusion of the relativistic spin-orbit coupling would further increase the number of molecular electronic states and complexity of the system. The entrance $B^1\Sigma^+$ electronic state is separated from higher lying electronic states, but its Al$^+$($^1S$)+Sr($^1S$) atomic threshold is just 2346$\,$cm$^{-1}$ above the lowest Al($^2P$)+Sr$^+$($^2S$) atomic threshold. This suggests that non-radiative charge transfer driven by the non-adiabatic radial couplings can compete with the radiative charge transfer in this case. In the case of the (Al+Sr)$^+$ system, as shown in Fig.~\ref{fig:dm_SrAl+}, the transition moment between the $B^1\Sigma^+$ and $X^1\Sigma^+$ electronic states is also noticeably larger than the one between the $B^1\Sigma^+$ and $A^1\Pi$ states, in favor of radiative processes leading to the $X^1\Sigma^+$ state. The equilibrium distance of the $B^1\Sigma^+$ state is much larger as compared to the $X^1\Sigma^+$ and $A^1\Pi$ states, hence a small overlap between vibrational states associated with these electronic states may be expected. Excited electronic states have equalibrium distances and well depths in a broad range of values.

The long-range electrostatic, induction, and dispersion interaction coefficients  describing the long-range part of the interaction potential energy curves between the Al$^+$ ion and the Rb or Sr atom and between the Rb$^+$ or Sr$^+$ ion and the Al atom are collected in Table~\ref{tab:Cn}. They are based on the electronic properties of monomers calculated in the present study: the static electric dipole and quadruple polarizabilities of the Rb and Sr atoms $\bar{\alpha}_\text{Rb}=319.5\,$a.u., $\bar{\alpha}_\text{Sr}=199.2\,$a.u., $\bar{\beta}_\text{Rb}=6578\,$a.u., $\bar{\beta}_\text{Sr}=4551\,$a.u., the isotropic and anisotropic components of the electric dipole and quadruple polarizabilities of the Al atom $\bar{\alpha}_\text{Al}=57.7\,$a.u., $\Delta{\alpha}_\text{Al}=25.4\,$a.u., $\bar{\beta}_\text{Al}=622\,$a.u., $\Delta\beta_\text{Al}=309\,$a.u., the electric quadruple moment of the Al atom $\Theta_\text{Al}=5.2\,$a.u., and the static electric dipole polarizabilities of the Al$^+$, Rb$^+$, and Sr$^+$ ions, $\bar{\alpha}_{\text{Al}^+}=24.2\,$a.u., $\bar{\alpha}_{\text{Rb}^+}=9.2\,$a.u., $\bar{\alpha}_{\text{Sr}^+}=92.0\,$a.u. These values agree well with previous measurements and calculations~\cite{MitroyJPB10}. 

An additional check of the accuracy of the employed \textit{ab initio} methods is a comparison of the ionization potentials and atomic excitation energies with experimental values. The calculated ionization potentials of the Al, Rb, and Sr atoms are 48123$\,$cm$^{-1}$, 33566$\,$cm$^{-1}$, and 45814$\,$cm$^{-1}$. They agree within 0.5\% with experimental values of 48278$\,$cm$^{-1}$, 33691$\,$cm$^{-1}$, and 45932$\,$cm$^{-1}$~\cite{nist}. The lowest non-relativistic excitation energies of the Al, Rb, Sr atoms are 25388$\,$cm$^{-1}$, 12686$\,$cm$^{-1}$, and 14639$\,$cm$^{-1}$. They also agree within 0.5\% with corresponding experimental values of 25273$\,$cm$^{-1}$, 12737$\,$cm$^{-1}$ and 14705$\,$cm$^{-1}$~\cite{nist}. The lowest computed excitation energy for the Sr$^+$ ion is 15370$\,$cm$^{-1}$ as compered to experimental value of 14724$\,$cm$^{-1}$.   

Finally, the uncertainty of the calculated potential energy curves and electronic properties is of the order of 5-15\% depending on the electronic state. This estimation is based on the analysis of the convergence with the size of the used atomic basis sets and the level of employed theory. The lack of the exact treatment of the triple and higher excitations in the employed CCSD(T) and MRCISD methods, infeasible computationally to include for the studied systems, is a primary limiting factor. The presented CCSD(T) and MRCISD results agree with each other within about 10$\,$\%.

\begin{figure}[tb]
\begin{center}
\includegraphics[width=\columnwidth]{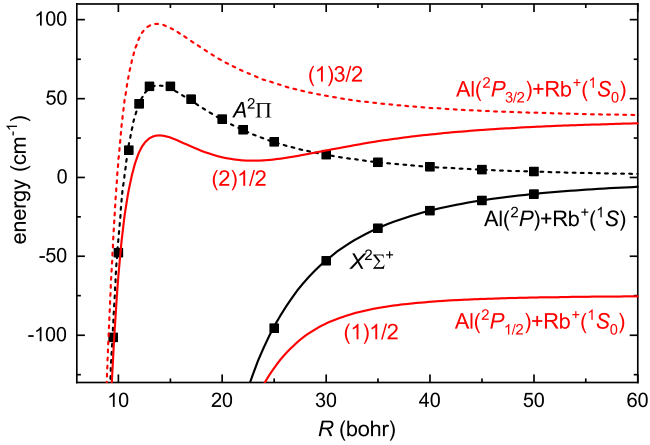}
\end{center}
\caption{The long-range part of the interaction potentials for the (Al+Rb)$^+$ ion-atom system. Points are the result of the supermolecular calculations and lines are the multipole expansion $-C_3/R^3-C_4/R^4-C_6/R^6$ with coefficients obtained within the perturbation theory. Black and red (gray) lines present non-relativistic and relativistic potential energy curves, respectively.}
\label{fig:long_range}
\end{figure}

The long-range part of the interaction potential is especially important for cold and ultracold collisions, when shape resonances, and quantum reflection and tunneling play a role. Figure~\ref{fig:long_range} presents the long-range part of the interaction potentials for the lowest asymptote of the (Al+Rb)$^+$ ion-atom system obtained in the molecular non-relativistic calculations together with the interaction potentials given by the multipole expansion of Eq.~\eqref{eq:long-range} obtained within the perturbation theory. Both approaches agree well in a broad range of interatomic distances. This agreement additionally validates the employed electronic structure methods and results from the size consistency of the coupled cluster method combined with the same level of theory used to describe dimer and monomers. Similar agreement is achieved for the (Al+Sr)$^+$ ion-atom system.  Additionally, the impact of the spin-orbit coupling in the (Al+Rb)$^+$ system is presented by showing the relativistic potential energy curves connected to two fine-state manifolds obtained by the diagonalization of Eqs.~\eqref{eq:12_RbAl+} and~\eqref{eq:32_RbAl+}.

The leading long-range induction interaction given by the $C_4$  coefficient for the Al$^+$ ion interacting with the Rb or Sr atom determines the characteristic interaction length scale $R_4 = \sqrt{{2\mu C_4}/{\hbar^2}}$ and the related energy scale $E_4 = {\hbar^2}/{2\mu R_4^2}$
relevant for ultracold ion-atom collisions~\cite{TomzaRMP18}. For the Al$^+$+Rb and Al$^+$+Sr systems they take values of $R_4$=3454$\,$bohr, $E_4$=355$\,$nK and $R_4$=2738$\,$bohr, $E_4$=560$\,$nK, respectively.

\subsection{Cold collisions and charge transfer dynamics}

In the considered ion-atom systems there are two paths of collision- and interaction-induced radiative charge rearrangement: (i) the radiative charge transfer (RCT)
\begin{equation}
\text{Al}^+ + X \to \text{Al} + X^+ + hv\,,
\end{equation}
where the electron is spontaneously transferred from the $X$=Rb(Sr) atom to the Al$^+$ ion emitting a photon of energy $hv$, and (ii) the radiative association (RA)
\begin{equation}
\text{Al}^+ + X \to \text{Al}X^+ + hv\,,
\end{equation}
where the Al$^+$ ion and the $X$=Rb(Sr) atom spontaneously form an Al$X^+$ molecular ion emitting a photon of energy~$hv$. Scattering of the Al$^+$ ion with the Rb(Sr) atom is governed by one potential energy curve of the $B^2\Sigma^+$($B^1\Sigma^+$) symmetry, whereas the charge transfer driven by the transition electric dipole moment can lead to two electronic states of the  $X^2\Sigma^+$ and $A^2\Pi$ ($X^1\Sigma^+$ and $A^1\Pi$) symmetries, respectively.

\begin{figure}[tb]
\begin{center}
\includegraphics[width=\columnwidth]{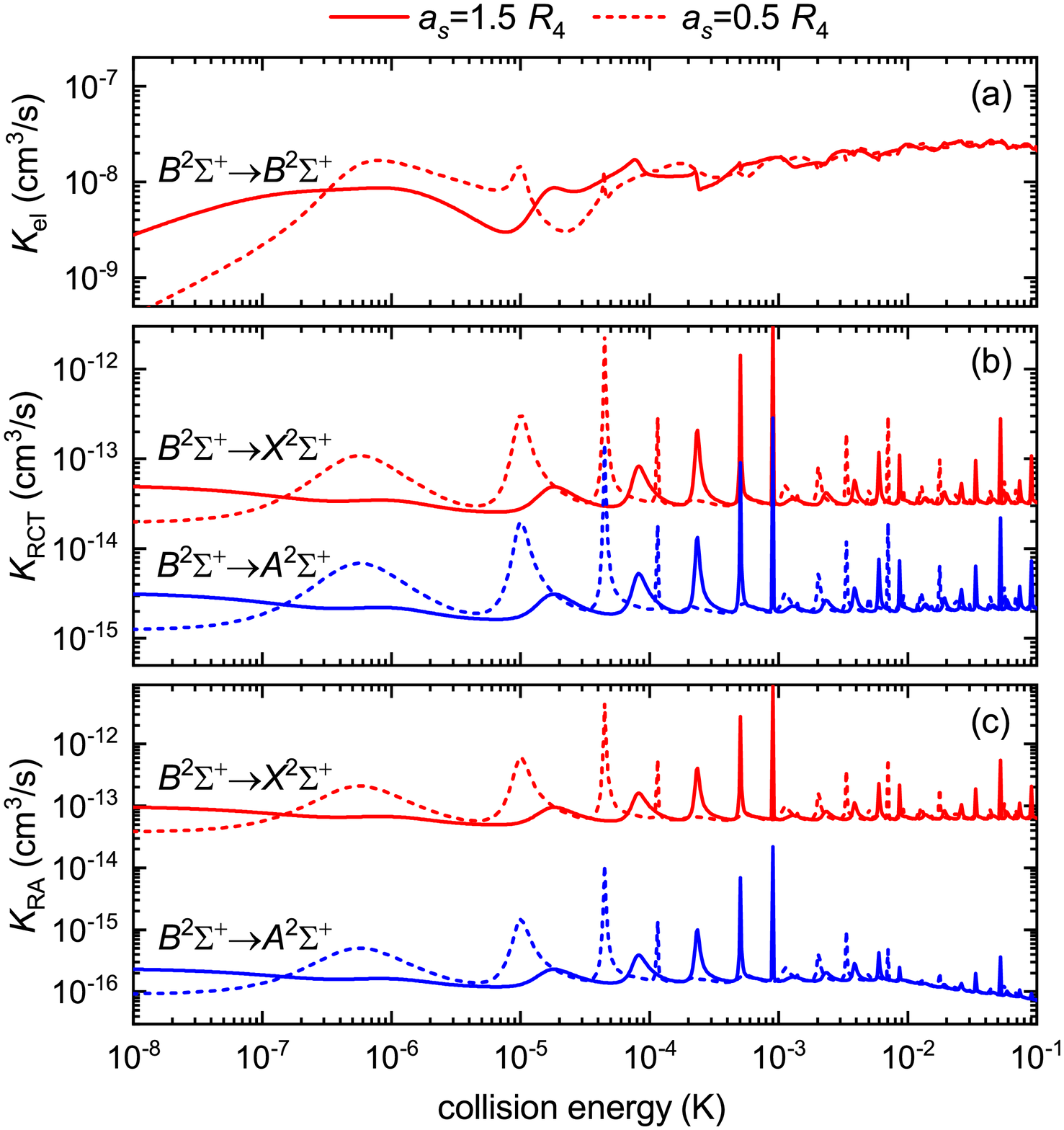}
\end{center}
\caption{Rate constants for collisions between the Al$^+$ ion and Rb atom in the $B^2\Sigma^+$ electronic state as a function of the collision energy: (a) elastic scattering, (b) radiative charge transfer, (c) radiative association. Reactive rate constants leading to the $X^2\Sigma^+$ and $A^2\Pi$ electronic states are shown separately. Results for scattering length set to $0.5R_4$ and $1.5R_4$ are presented.
}
\label{fig:K_AlRb+}
\end{figure}

Figure~\ref{fig:K_AlRb+} presents rate constants for elastic and inelastic reactive collisions between the Al$^+$ ion and Rb atom as a function of the collision energy. Results for the Al$^+$ ion and Sr atom have similar energy dependence but the magnitude of rate constants for inelastic collisions is different. Similarly to other ion-atom systems~\cite{KrychPRA11,IdziaszekNJP11,SayfutyarovaPRA13,TomzaPRA15a,daSilvaNJP2015}, shape resonances are much more pronounced for inelastic rate constants, however if the thermal Maxweell-Boltzman or Tsallis distribution of collision energies is assumed~\cite{RousePRL17}, the thermal averaging removes energy dependence for temperatures larger than 1$\,$mK in agreement with predictions of the classical Langevin capture theory~\cite{Levine09}. Scattering results are presented for two typical scattering lengths ($a_s$=0.5$\,R_4$ and $a_s$=1.5$\,R_4$) in the entrance channel. Rate constants for small collision energies and patter of shape resonances depend strongly on theses scattering lengths, but the overall magnitude of rate constants does not depend on them. Additionally, all presented results does not depend on the scattering lengths in the exit channels. 

In the range of investigated collision energies, the rate constants for the elastic scattering $K_\text{el}$ for both ion-atom systems have similar values of (see Fig.~\ref{fig:K_AlRb+}(a))
$$K_\text{el}(B^2\Sigma^+)\approx K_\text{el}({B^1\Sigma^+}) \approx 10^{-8}\,\text{cm}^3/\text{s}\,.$$
For comparison, the inelastic rate constants given by the classical Langevin capture model determined by the long-range $C_4$ coefficients $K_\text{L}=2\pi\sqrt{2 C_4/\mu}$~\cite{Levine09} take values of $K_\text{L}=3.6\cdot 10^{-9}\,\text{cm}^3/\text{s}$ and $K_\text{L}=2.8\cdot 10^{-9}\,\text{cm}^3/\text{s}$ for the Al$^+$+Rb and Al$^+$+Sr systems, respectively.

The rate constants for the inelastic scattering depend strongly on the system and exit electronic state. 
The rate constants for the radiative charge transfer $K_\text{RCT}$ for the Al$^+$ ion colliding with the Rb atom are (see Fig.~\ref{fig:K_AlRb+}(b))
$$K_\text{RCT}(B^2\Sigma^+\to X^2\Sigma^+)\approx 4.5\cdot 10^{-14}\,\text{cm}^3/\text{s}\,, $$
$$K_\text{RCT}(B^2\Sigma^+\to A^2\Pi)\approx 3\cdot 10^{-15}\,\text{cm}^3/\text{s}\,, $$
whereas for the Al$^+$ ion colliding with the Sr atom they are
$$K_\text{RCT}(B^1\Sigma^+\to X^1\Sigma^+)\approx 1.4 \cdot 10^{-17}\,\text{cm}^3/\text{s}\,, $$
$$K_\text{RCT}(B^1\Sigma^+\to A^1\Pi)\approx 5 \cdot 10^{-19}\,\text{cm}^3/\text{s}\,. $$
Interestingly, the probability of the charge transfer to the $X^2\Sigma^+$ state is 15 times larger than to the $A^2\Pi$ one in Al$^+$+Rb collisions and the probability of the charge transfer to the $X^1\Sigma^+$ state is 25 times larger than to the $A^1\Pi$ one in Al$^+$+Sr collisions. The main reason for observed branching rations is the difference in electric transition dipole moments driving the radiative charge transfer between considered electronic states, which is few times larger for $B\to X$ transitions as compared to $B\to A$ ones (see Figs.~\ref{fig:dm_RbAl+} and~\ref{fig:dm_SrAl+}). The rate constants for spontaneous radiative transitions are proportional to the square of transition dipole moments. Additionally, the probability of the $B\to X$ and $B\to A$ charge transfer for Al$^+$+Rb collisions is 6000 and 3000 times larger than for Al$^+$+Sr collisions, respectively. The main reason for this difference is much smaller energy of emitted photons (the difference of the ionization potentials) in Al$^+$+Sr collisions as compared to Al$^+$+Rb ones (2346.4$\,$cm$^{-1}$ vs.~14587.7$\,$cm$^{-1}$~\cite{nist}). The rate constants for spontaneous radiative transitions are proportional to the cubic of the transition energy. The second reason is a larger misalignment of potential wells of the entrance and exit electronic states in the (Al+Sr)$^+$ system as compared to the (Al+Rb)$^+$ one. The rate constants for the non-radiative charge transfer driven by the non-adiabatic radial coupling in Al$^+$+Rb (Al$^+$+Sr) collisions estimated using Landau-Zener formula~\cite{BelyaevPRA11} are at least ten orders (one order) of magnitude smaller than the respective radiative rate constants, in agreement with calculations for other systems~\cite{RellergertPRL11,SayfutyarovaPRA13}. Therefore we neglect them in this paper.

The rate constants for the radiative association $K_\text{RA}$ for the Al$^+$ ion colliding with the Rb atom are (see Fig.~\ref{fig:K_AlRb+}(c))
$$K_\text{RA}(B^2\Sigma^+\to X^2\Sigma^+)\approx 9 \cdot 10^{-14}\,\text{cm}^3/\text{s}\,,$$
$$K_\text{RA}(B^2\Sigma^+\to A^2\Pi)\approx 2 \cdot 10^{-16}\,\text{cm}^3/\text{s}\,,$$
whereas for the Al$^+$ ion colliding with the Sr atom they are
$$K_\text{RA}(B^1\Sigma^+\to X^1\Sigma^+)\approx 3.5 \cdot 10^{-15}\,\text{cm}^3/\text{s}\,,$$
$$K_\text{RA}(B^1\Sigma^+\to A^1\Pi)\approx 4 \cdot 10^{-31}\,\text{cm}^3/\text{s}\,.$$
Interestingly, the probability of the radiative association to the $X^2\Sigma^+$ state is 450 times larger than to the $A^2\Pi$ one in Al$^+$+Rb collisions and the radiative association to the $A^1\Pi$ state is strongly suppressed in Al$^+$+Sr collisions. Less favorable transition electric dipole moments for the $B\to A$ association than for the $B\to X$ one are partially responsible for the observed suppression of the rate constants, but the main reason for it is the repulsive character of the $A^2\Pi$ and $A^1\Pi$ electronic states at intermediate and large distances due to the strongly repulsive nature of the ion-quadruple interaction for $\Pi$ electronic states and associated small potential well depths (see Fig.~\ref{fig:long_range}). Consequently, these electronic states support much smaller number and density of ro-vibrational states as compared to $\Sigma$ ones (around 10 vibrational levels for $A^2\Pi$ and $A^1\Pi$ to be compared to around 200 vibrational levels for $X^2\Sigma^+$ and $X^1\Sigma^+$). Additionally, the short-range classical turning point for the $B^1\Sigma^+$ state in the (Al+Sr)$^+$ system overlap with the maximum of the electrostatic barrier of the $A^1\Pi$ state, thus resulting in highly unfavorable Franck-Condon factors.  The probability of the $B\to X$ radiative association for Al$^+$+Rb collisions is only 25 times larger than for Al$^+$+Sr collisions. For collision energies larger than 10$\,$mK, the rate constants for the  $B\to A$ radiative association in both systems start to decrease with increasing collision energy, because the centrifugal barrier in the $A$ electronic states starts to be comparable to the electronic barrier and depth of these states, and suppresses the number of bound ro-vibrational levels which can be populated. 

For alkaline-earth-metal ions colliding with alkali-metal atoms, typically, the rate constants for the radiative association are predicted to be significantly larger than the rate constants for the radiative charge transfer~\cite{TomzaPRA15a,daSilvaNJP2015}. In the present case, because of the interplay of the short-range and long-range effects described in the previous paragraphs, the patter is more complex. For the $B\to X$ transition, the radiative association is twice more probable than the radiative charge transfer for Al$^+$+Rb collisions but the radiative charge transfer is 250 times more probable than the radiative association for Al$^+$+Sr collisions. For both systems, the $B\to A$ transitions are less probably than the $B\to X$ ones. For $B\to A$ transition, the radiative charger transfer is 13.5 times more probable than the radiative association for Al$^+$+Rb collisions and several orders of magnitude more probable for Al$^+$+Sr collisions.

In the non-relativistic description of electronic symmetries, the radiative transitions $K_\text{R}$ ($K_\text{RA}$ or $K_\text{RCT}$) to the triplet $a^3\Pi$ and $b^3\Sigma^+$ electronic states in Al$^+$+Sr collisions are strictly electric dipole forbidden
$$K_\text{R}(B^1\Sigma^+\to a^3\Pi)=K_\text{R}(B^1\Sigma^+\to b^3\Sigma^+)=0\,.$$
The inclusion of the spin-orbit coupling, which is the leading relativistic effect and mixes different non-relativistic electronic states, slightly modifies the pattern of inelastic rate constants described above.  

In Al$^+$+Rb collisions, the rate constants for the radiative charge transfer and association leading to the $|\Omega|=3/2$ state are almost the same as they are for transitions leading to the $A^2\Pi$ state
$$K_\text{R}((3)1/2\to (1)3/2)\approx  K_\text{R}(B^2\Sigma^+\to A^2\Pi)\,, $$
with a small modification due to the change of emitted photons' energy. The relativistic $|\Omega|=1/2$ electronic states are combinations of the non-relativistic $X^2\Sigma^+$ and $A^2\Pi$ states, however the significant mixing occurs only close to the dissociation threshold whereas the interaction wells of the $(1)1/2$ and $(2)1/2$ states are mostly of the $X^2\Sigma^+$ and $A^2\Pi$ nature, respectively. The rate constants for the radiative charge transfer and association to the $(1)1/2$ state are within a few percent as compared to the ones to the $X^2\Sigma^+$ state in the non-relativistic picture  
$$K_\text{R}((3)1/2\to (1)1/2)\approx K_\text{R}(B^2\Sigma^+\to X^2\Sigma^+)\,.$$
The radiative transitions to the $(2)1/2$ state are larger as compared to the ones to the $A^2\Pi$ state 
$$K_\text{RCT}((3)1/2\to (2)1/2)\approx 2 \cdot K_\text{RCT}(B^2\Sigma^+\to A^2\Pi)\,, $$
$$K_\text{RA}((3)1/2\to (2)1/2)\approx 3 \cdot K_\text{RA}(B^2\Sigma^+\to A^2\Pi)\,,$$
because the spin-orbit coupling and mixing of the $X^2\Sigma^+$ and $A^2\Pi$ states remove the electronic barrier and repulsive character of the ion-quadruple interaction in the $(2)1/2$ electronic state (see~Fig.~\ref{fig:long_range}). 

The branching ratio of radiative charge transfer collisions leading to the Al atom in the $^2P_{1/2}$ and $^2P_{3/2}$ fine states is 
$$ \frac{K_\text{RCT}(\text{Al}^+ + \text{Rb}\to \text{Al}({}^2P_{1/2})+\text{Rb}^+)}{K_\text{RCT}(\text{Al}^+ + \text{Rb}  \to \text{Al}({}^2P_{3/2})+ \text{Rb}^+)}\approx 5\,.$$

In Al$^+$+Sr collisions, radiative transitions leading to the $|\Omega|=2$ state are still electric dipole forbidden
$$K_\text{R}((3)0^+\to (1)2)=0\,,$$
however the non-relativistic $a^3\Pi$ and $b^3\Sigma^+$ electronic states contribute to the relativistic $\Omega=0^+$ and $|\Omega|=1$ states.

The relativistic $\Omega=0^+$ electronic states are combinations of the non-relativistic $X^1\Sigma^+$ and $a^3\Pi$ states, but the significant mixing occurs only close to the dissociation threshold whereas the interaction wells of the $(1)0^+$ and $(2)0^+$ states are mostly of the $X^1\Sigma^+$ and $a^3\Pi$ nature, respectively. The rate constants for the radiative charge transfer to the $(1)0^+$ state is 10 times smaller as compared to the ones to the $X^1\Sigma^+$ state in the non-relativistic picture  
$$K_\text{RCT}((3)0^+\to (1)0^+)\approx 0.1\cdot K_\text{RCT}(B^1\Sigma^+\to X^1\Sigma^+)\,, $$
while the rate constants for the radiative association to the $(1)0^+$ state are just a few percent smaller as compared to the ones to the $X^1\Sigma^+$ state
$$K_\text{RA}((3)0^+\to (1)0^+)\approx K_\text{RA}(B^1\Sigma^+\to X^1\Sigma^+)\,.$$
The radiative transitions to the $(2)0^+$ state which is mostly of the $a^3\Pi$ nature are possible because of the admixture of the $X^1\Sigma^+$ state 
$$K_\text{RCT}((3)0^+\to (2)0^+)\approx 0.2 \cdot K_\text{RCT}(B^1\Sigma^+\to X^1\Sigma^+)\,, $$
$$K_\text{RA}((3)0^+\to (2)0^+)\approx 0.00005 \cdot K_\text{RA}(B^1\Sigma^+\to X^1\Sigma^+)\,.$$
The radiative charge transfer is more affected by the spin-orbit coupling than the radiative association in this case because it is governed by transitions at larger distances where mixing of electronic states and transition moments is larger.

The relativistic $|\Omega|=1$ electronic states are combinations of the non-relativistic $A^1\Pi$, $a^3\Pi$, and $b^3\Sigma^+$ states, where as in previous cases the spin-orbit coupling significantly mixes states close to the dissociation threshold and thus affects more the charge transfer. Additionally, the wells of the $(1)1$ and $(2)1$ states are mixtures of the crossing $a^3\Pi$ and $b^3\Sigma^+$ states, whereas the interaction well of the $(3)1$ state is mostly of the $A^1\Pi$ nature. The resulting rate constants for the radiative charge transfer are
$$K_\text{RCT}((3)0^+\to (1)1)\approx 0.002\cdot K_\text{RCT}(B^1\Sigma^+\to A^1\Pi)\,, $$
$$K_\text{RCT}((3)0^+\to (2)1)\approx 0.15\cdot K_\text{RCT}(B^1\Sigma^+\to A^1\Pi)\,, $$
$$K_\text{RCT}((3)0^+\to (3)1)\approx 0.8\cdot K_\text{RCT}(B^1\Sigma^+\to A^1\Pi)\,, $$
while the rate constants for the radiative association are
$$K_\text{RA}((3)0^+\to (1)1)\approx 2.2\cdot K_\text{RCT}((3)0^+\to (1)1)\,, $$
$$K_\text{RA}((3)0^+\to (2)1)\approx 0.2\cdot K_\text{RCT}((3)0^+\to (2)1)\,, $$
$$K_\text{RA}((3)0^+\to (3)1)\approx 10^{-8}\cdot K_\text{RCT}((3)0^+\to (3)1)\,. $$

The branching ratio of radiative charge transfer collisions leading to the Al atom in the $^2P_{1/2}$ and $^2P_{3/2}$ fine states is 
$$ \frac{K_\text{RCT}(\text{Al}^+ + \text{Sr}\to \text{Al}({}^2P_{1/2})+\text{Sr}^+)}{K_\text{RCT}(\text{Al}^+ + \text{Sr}  \to \text{Al}({}^2P_{3/2})+ \text{Sr}^+)}\approx \frac{1}{2}\,.$$

\begin{figure}[tb]
\begin{center}
\includegraphics[width=\columnwidth]{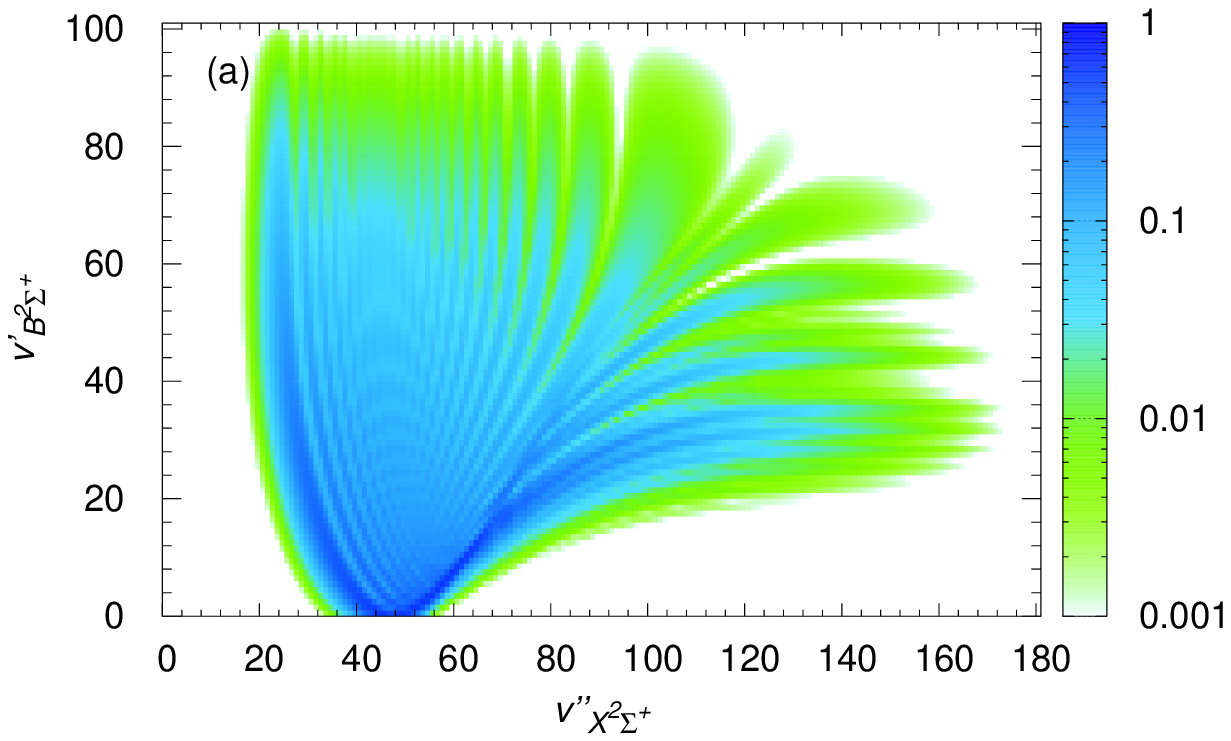}
\includegraphics[width=\columnwidth]{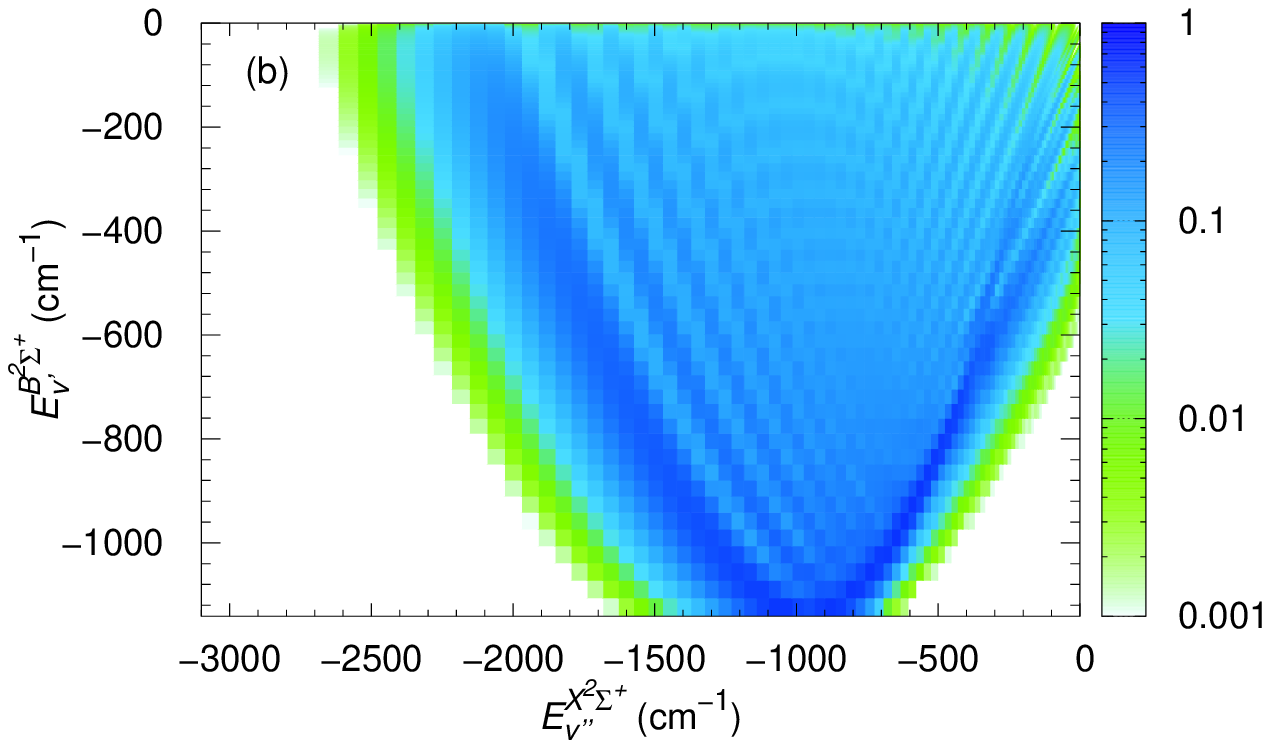}
\end{center}
\caption{Transition electric dipole moments between vibrational levels of the $X^2\Sigma^+$ ground and $B^2\Sigma^+$ excited electronic states of the (Al+Rb)$^+$ ion-atom system as a function of the vibrational quantum numbers (a) and vibrational energies (b).}
\label{fig:dvv}
\end{figure}

Finally, all calculated rate constants for the radiative charge transfer and association are significantly smaller (at least $10^4$ times smaller) than the Langevin rate constants. This means that at least $10^4$ collisions is needed to observe reactive processes in the Al$^+$+Rb and Al$^+$+Sr systems. In typical hybrid ion-atom experiments measuring such small rate constants is challenging because long interrogation times are needed. The employment of the precision measurement within the generalized quantum logic spectroscopy scheme, where an alkaline-earth ion is used to monitor the replacement of the Al$^+$ ion by Rb$^+$(Sr$^+$) or AlRb$^+$(AlSr$^+$) and their energy,  may allow to assess statistics of these relatively rare events due to enhanced controllability and sensitivity of such a scheme~\cite{Brewer2019b,Guggemos2019}. On the other hand, small reactivity can be an advantage providing long interrogation times to probe even weak ion-atom interactions. The measurement of the charge and energy transfer on the single-collision level can also be envisioned~\cite{MeirPRL16,MeirPRL18}. In fact, sympathetic cooling in the considered ion-atom systems may be challenging due to the micromotion-induced heating~\cite{CetinaPRL12}. On the other hand, recent developments on optical ion trapping may be a remedy~\cite{Schmidt2019,SchneiderNatPhot10}.

\begin{figure}[tb]
\begin{center}
\includegraphics[width=\columnwidth]{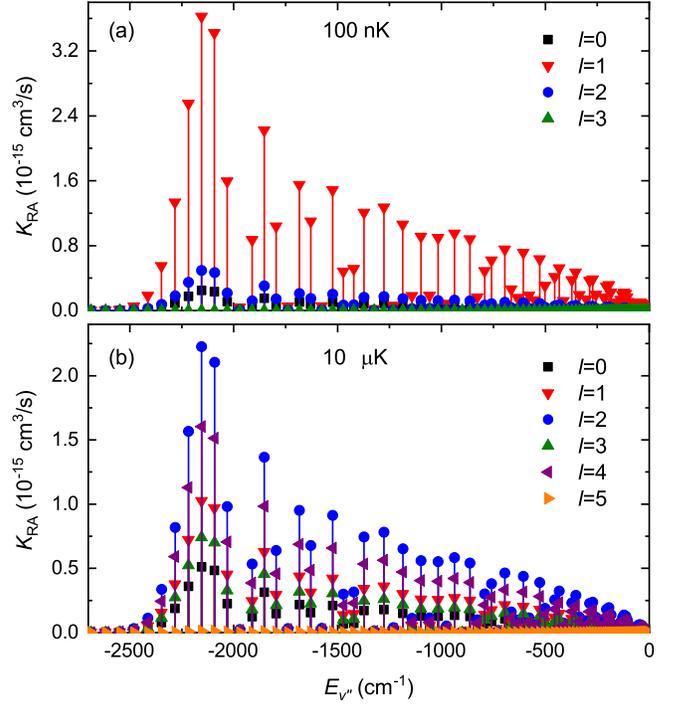}
\end{center}
\caption{Radiative association rate constants as a function of the energy of the final ro-vibrational level in the $X^2\Sigma^+$ electronic ground state of the AlRb$^+$ molecular ion for the Al$^+$ ion colliding with the Rb atom in the $B^2\Sigma^+$ state with the scattering length of $R_4$ at the temperature of 100$\,$nK~(a) and 10$\,\mu$K~(b).}
\label{fig:Kv}
\end{figure}

\subsection{Formation and control of molecular ions}

In the previous subsection, we have analyzed the branching ratios of the radiative association to different electronic states and showed that formation of molecular ions is more probable in the $X^2\Sigma^+$ and $X^1\Sigma^+$ (and related relativistic) states than in the $A^2\Sigma^+$ and $A^1\Sigma^+$ ones in Al$^+$+Rb and Al$^+$+Sr collisions, respectively. Now, we investigate the branching ratios to different vibrational levels of the ground electronic states.
Figure~\ref{fig:dvv} presents the transition electric dipole moments between vibrational levels of the $X^2\Sigma^+$ ground and $B^2\Sigma^+$ excited electronic states of the (Al+Rb)$^+$ ion-atom system as a function of the vibrational quantum numbers in the upper panel and vibrational energies in the bottom panel. These transition dipole moments govern both spontaneous and stimulated transitions between different vibrational levels. The presented map of the transition electric dipole moments can be useful to engineer molecular ions in selected vibrational levels. Similar calculations were realized for the (Al+Sr)$^+$ ion-atom system and other electronic states in both non-relativistic and relativistic pictures. Unfortunately, there is no efficient path for the formation of the absolute ground-state molecular ions using the B and X electronic states.

Radiative association rate constants as a function of the energy of the final ro-vibrational level in the $X^2\Sigma^+$ electronic ground state for the Al$^+$ ion colliding with the Rb atom in the $B^2\Sigma^+$ state are plotted in Fig.~\ref{fig:Kv}. In Al$^+$+Rb collisions, the formation of molecular ions in vibrational levels with the vibrational quantum number around $v=24$ and the binding energy around 2155$\,$cm$^{-1}$ of the $X^2\Sigma^+$ electronic state is the most probable. In Al$^+$+Sr collisions, the formation of molecular ions in vibrational levels with the vibrational quantum number around $v=88$ and the binding energy around 1709$\,$cm$^{-1}$ of the $X^1\Sigma^+$ electronic state is the most probable. In both systems, the molecular formation probability decreases gradually for decreasing binding energies and is strongly suppressed for binding energies larger than 2500$\,$cm$^{-1}$. In both systems, the formation of molecular ions in the $A$ electronic states is the most probable in the last and most-weakly-bound vibrational level due to the suppression described in the previous subsection. Collisions at temperature of 100$\,$nK in Fig.~\ref{fig:Kv}(a) are dominated by $s$ wave, therefore, molecular ions in the first excited rotational state ($l$=1) are mostly formed. At temperature of 10$\,\mu$K in Fig.~\ref{fig:Kv}(b) several partial waves already contribute in collisions, therefore, broader range of rotational molecular states is populated.

In all considered cases of the radiative association, the formation of deeply-bound molecular ions close the the ground vibrational level $v=0$ is strongly suppressed due to the misalignment of the potential wells of the entrance B and exit X and A electronic states. To overcome this problem and to enable the formation of ground-state molecular ions with larger binding energies, the two-photon schemes using an intermediate electronic state can be employed~\cite{GacesaPRA16}. In such scenarios, colliding ion-atom pairs can be excited to a selected intermediate excited electronic state, which has better overlaps with both the initial B and final X and A states. Next, spontaneous or stimulated emission can result in the formation of deeply-bound molecular ions. For the (Al+Rb)$^+$ and (Al+Sr)$^+$ ion-atom systems, the $D^2\Pi$ and $(3)^1\Pi$ excited electronic state can be used as intermediate states in the two-photon formation, respectively. After the initial spontaneous or stimulated radiative association, the stabilization to the ground vibrational level can also be realized with two-photon schemes using Stimulated Raman adiabatic passage (STIRAP)~\cite{VitanovRMP17}. The quantum logic spectroscopy can also be used for state-selective detection of formed molecular ions based on different rotational consonants and dipole moments of molecular ions in different vibrational states~\cite{KuangPRA11,WolfNature16}.

\section{Summary and conclusions}
\label{sec:summary}

Motivated by recent experimental advances on highly precise atomic clocks based on sympathetically cooled Al$^+$ ions, on one hand, and on the preparation and application of cold ion-atom systems, on the other hand, here we have proposed to employ an Al$^+$ ion co-trapped in a Paul trap with a laser-cooled alkaline-earth ion and overlapped with a small cloud of ulracold Rb or Sr atoms to probe ulracold ion-atom interactions and collisions, including radiative charge exchange and association processes.

We have calculated potential energy curves and transition electric dipole moments for the prototype (Al+Rb)$^+$ and (Al+Sr)$^+$ ion-atom systems using state-of-the-art \textit{ab initio} electronic structure techniques of quantum chemistry such as the coupled cluster and multireference configuration interaction methods with scalar relativistic effects included within the small-core energy-consistent pseudopotentials. We have also used perturbation theory to obtain the long-range interaction coefficients and to include the relativistic spin-orbit coupling between non-relativistic molecular electronic states. Next, the electronic structure data have been employed to investigate cold collisions and charge transfer dynamics.

Scattering of an Al$^+$ ion with alkali-metal or alkaline-earth-metal atom is governed by a singe potential energy curve whereas dipole allowed transitions can lead to several electronic states mixed by the spin-orbit coupling. Such transitions can result in the radiative charge transfer or radiative association. We have analyzed the interplay of the short-range and long-range effects, and examined the branching ratios for those rich and interesting processes. We have also investigated the prospects for the laser-field control and formation of molecular ions. 

The atomic clock transition in an Al$^+$ ion can be employed to monitor ion-atom interactions and scattering dynamics \textit{via} quantum logic spectroscopy~\cite{SchmidtScience05}. In this method, the quantum state detection and preparation of the Al$^+$ ion can be realized by optical addressing of a co-trapped alkaline-earth ion. The frequency shift of trapped-ion optical clocks caused by background-gas collisions have recently been actively studied~\cite{GibblePRL13,VuthaPRA17,Vutha2018,Hankin2019,Davis2019}. High precision measurement of the narrow clock transition may allow to probe even weak ion-atom interactions and long interrogation times may allow to assess statistics of relatively rare events of charge-transfer collisions~\cite{Brewer2019b,Guggemos2019}. The charge and energy transfer may also potentially be investigated on the single-collision level~\cite{MeirPRL18}.  In summary, the presented results pave the way for the application of ions other than alkali-metal and alkaline-earth-metal ones in the field of cold hybrid ion-atom experiments.

\begin{acknowledgments}
Financial support from the National Science Centre Poland (2016/23/B/ST4/03231) and the PL-Grid Infrastructure is gratefully acknowledged. 
\end{acknowledgments}

\bibliography{XAl+}

\begin{thebibliography}{92}%
\makeatletter
\providecommand \@ifxundefined [1]{%
 \@ifx{#1\undefined}
}%
\providecommand \@ifnum [1]{%
 \ifnum #1\expandafter \@firstoftwo
 \else \expandafter \@secondoftwo
 \fi
}%
\providecommand \@ifx [1]{%
 \ifx #1\expandafter \@firstoftwo
 \else \expandafter \@secondoftwo
 \fi
}%
\providecommand \natexlab [1]{#1}%
\providecommand \enquote  [1]{``#1''}%
\providecommand \bibnamefont  [1]{#1}%
\providecommand \bibfnamefont [1]{#1}%
\providecommand \citenamefont [1]{#1}%
\providecommand \href@noop [0]{\@secondoftwo}%
\providecommand \href [0]{\begingroup \@sanitize@url \@href}%
\providecommand \@href[1]{\@@startlink{#1}\@@href}%
\providecommand \@@href[1]{\endgroup#1\@@endlink}%
\providecommand \@sanitize@url [0]{\catcode `\\12\catcode `\$12\catcode
  `\&12\catcode `\#12\catcode `\^12\catcode `\_12\catcode `\%12\relax}%
\providecommand \@@startlink[1]{}%
\providecommand \@@endlink[0]{}%
\providecommand \url  [0]{\begingroup\@sanitize@url \@url }%
\providecommand \@url [1]{\endgroup\@href {#1}{\urlprefix }}%
\providecommand \urlprefix  [0]{URL }%
\providecommand \Eprint [0]{\href }%
\providecommand \doibase [0]{http://dx.doi.org/}%
\providecommand \selectlanguage [0]{\@gobble}%
\providecommand \bibinfo  [0]{\@secondoftwo}%
\providecommand \bibfield  [0]{\@secondoftwo}%
\providecommand \translation [1]{[#1]}%
\providecommand \BibitemOpen [0]{}%
\providecommand \bibitemStop [0]{}%
\providecommand \bibitemNoStop [0]{.\EOS\space}%
\providecommand \EOS [0]{\spacefactor3000\relax}%
\providecommand \BibitemShut  [1]{\csname bibitem#1\endcsname}%
\let\auto@bib@innerbib\@empty
\bibitem [{\citenamefont {Wineland}(2013)}]{WinelandRMP13}%
  \BibitemOpen
  \bibfield  {author} {\bibinfo {author} {\bibfnamefont {D.~J.}\ \bibnamefont
  {Wineland}},\ }\href {\doibase 10.1103/RevModPhys.85.1103} {\bibfield
  {journal} {\bibinfo  {journal} {Rev. Mod. Phys.}\ }\textbf {\bibinfo {volume}
  {85}},\ \bibinfo {pages} {1103} (\bibinfo {year} {2013})}\BibitemShut
  {NoStop}%
\bibitem [{\citenamefont {Haeffner}\ \emph {et~al.}(2008)\citenamefont
  {Haeffner}, \citenamefont {Roos},\ and\ \citenamefont
  {Blatt}}]{HaeffnerPR08}%
  \BibitemOpen
  \bibfield  {author} {\bibinfo {author} {\bibfnamefont {H.}~\bibnamefont
  {Haeffner}}, \bibinfo {author} {\bibfnamefont {C.~F.}\ \bibnamefont {Roos}},
  \ and\ \bibinfo {author} {\bibfnamefont {R.}~\bibnamefont {Blatt}},\ }\href
  {\doibase 10.1016/j.physrep.2008.09.003} {\bibfield  {journal} {\bibinfo
  {journal} {Phys. Rep.}\ }\textbf {\bibinfo {volume} {469}},\ \bibinfo {pages}
  {155} (\bibinfo {year} {2008})}\BibitemShut {NoStop}%
\bibitem [{\citenamefont {Blatt}\ and\ \citenamefont
  {Roos}(2012)}]{BlattNatPhys12}%
  \BibitemOpen
  \bibfield  {author} {\bibinfo {author} {\bibfnamefont {R.}~\bibnamefont
  {Blatt}}\ and\ \bibinfo {author} {\bibfnamefont {C.}~\bibnamefont {Roos}},\
  }\href {\doibase 10.1038/nphys2252} {\bibfield  {journal} {\bibinfo
  {journal} {Nat. Phys.}\ }\textbf {\bibinfo {volume} {8}},\ \bibinfo {pages}
  {277} (\bibinfo {year} {2012})}\BibitemShut {NoStop}%
\bibitem [{\citenamefont {Wineland}\ \emph {et~al.}(2002)\citenamefont
  {Wineland}, \citenamefont {Bergquist}, \citenamefont {Bollinger},
  \citenamefont {Drullinger},\ and\ \citenamefont {Itano}}]{Wineland02}%
  \BibitemOpen
  \bibfield  {author} {\bibinfo {author} {\bibfnamefont {D.~J.}\ \bibnamefont
  {Wineland}}, \bibinfo {author} {\bibfnamefont {J.~C.}\ \bibnamefont
  {Bergquist}}, \bibinfo {author} {\bibfnamefont {J.~J.}\ \bibnamefont
  {Bollinger}}, \bibinfo {author} {\bibfnamefont {R.~E.}\ \bibnamefont
  {Drullinger}}, \ and\ \bibinfo {author} {\bibfnamefont {W.~M.}\ \bibnamefont
  {Itano}},\ }\href@noop {} {\bibfield  {journal} {\bibinfo  {journal} {Proc.
  6th Symp. Frequency Standards and Metrology}\ } (\bibinfo {year}
  {2002})}\BibitemShut {NoStop}%
\bibitem [{\citenamefont {Degen}\ \emph {et~al.}(2017)\citenamefont {Degen},
  \citenamefont {Reinhard},\ and\ \citenamefont {Cappellaro}}]{DegenRMP17}%
  \BibitemOpen
  \bibfield  {author} {\bibinfo {author} {\bibfnamefont {C.~L.}\ \bibnamefont
  {Degen}}, \bibinfo {author} {\bibfnamefont {F.}~\bibnamefont {Reinhard}}, \
  and\ \bibinfo {author} {\bibfnamefont {P.}~\bibnamefont {Cappellaro}},\
  }\href {\doibase 10.1103/RevModPhys.89.035002} {\bibfield  {journal}
  {\bibinfo  {journal} {Rev. Mod. Phys.}\ }\textbf {\bibinfo {volume} {89}},\
  \bibinfo {pages} {035002} (\bibinfo {year} {2017})}\BibitemShut {NoStop}%
\bibitem [{\citenamefont {Kozlov}\ \emph {et~al.}(2018)\citenamefont {Kozlov},
  \citenamefont {Safronova}, \citenamefont {Crespo L\'opez-Urrutia},\ and\
  \citenamefont {Schmidt}}]{KozlovRMP18}%
  \BibitemOpen
  \bibfield  {author} {\bibinfo {author} {\bibfnamefont {M.~G.}\ \bibnamefont
  {Kozlov}}, \bibinfo {author} {\bibfnamefont {M.~S.}\ \bibnamefont
  {Safronova}}, \bibinfo {author} {\bibfnamefont {J.~R.}\ \bibnamefont {Crespo
  L\'opez-Urrutia}}, \ and\ \bibinfo {author} {\bibfnamefont {P.~O.}\
  \bibnamefont {Schmidt}},\ }\href {\doibase 10.1103/RevModPhys.90.045005}
  {\bibfield  {journal} {\bibinfo  {journal} {Rev. Mod. Phys.}\ }\textbf
  {\bibinfo {volume} {90}},\ \bibinfo {pages} {045005} (\bibinfo {year}
  {2018})}\BibitemShut {NoStop}%
\bibitem [{\citenamefont {Schmidt}\ \emph {et~al.}(2005)\citenamefont
  {Schmidt}, \citenamefont {Rosenband}, \citenamefont {Langer}, \citenamefont
  {Itano}, \citenamefont {Bergquist},\ and\ \citenamefont
  {Wineland}}]{SchmidtScience05}%
  \BibitemOpen
  \bibfield  {author} {\bibinfo {author} {\bibfnamefont {P.~O.}\ \bibnamefont
  {Schmidt}}, \bibinfo {author} {\bibfnamefont {T.}~\bibnamefont {Rosenband}},
  \bibinfo {author} {\bibfnamefont {C.}~\bibnamefont {Langer}}, \bibinfo
  {author} {\bibfnamefont {W.~M.}\ \bibnamefont {Itano}}, \bibinfo {author}
  {\bibfnamefont {J.~C.}\ \bibnamefont {Bergquist}}, \ and\ \bibinfo {author}
  {\bibfnamefont {D.~J.}\ \bibnamefont {Wineland}},\ }\href {\doibase
  10.1126/science.1114375} {\bibfield  {journal} {\bibinfo  {journal}
  {Science}\ }\textbf {\bibinfo {volume} {309}},\ \bibinfo {pages} {749}
  (\bibinfo {year} {2005})}\BibitemShut {NoStop}%
\bibitem [{\citenamefont {Rosenband}\ \emph {et~al.}(2008)\citenamefont
  {Rosenband}, \citenamefont {Hume}, \citenamefont {Schmidt}, \citenamefont
  {Chou}, \citenamefont {Brusch}, \citenamefont {Lorini}, \citenamefont
  {Oskay}, \citenamefont {Drullinger}, \citenamefont {Fortier}, \citenamefont
  {Stalnaker}, \citenamefont {Diddams}, \citenamefont {Swann}, \citenamefont
  {Newbury}, \citenamefont {Itano}, \citenamefont {Wineland},\ and\
  \citenamefont {Bergquist}}]{RosenbandScience08}%
  \BibitemOpen
  \bibfield  {author} {\bibinfo {author} {\bibfnamefont {T.}~\bibnamefont
  {Rosenband}}, \bibinfo {author} {\bibfnamefont {D.~B.}\ \bibnamefont {Hume}},
  \bibinfo {author} {\bibfnamefont {P.~O.}\ \bibnamefont {Schmidt}}, \bibinfo
  {author} {\bibfnamefont {C.~W.}\ \bibnamefont {Chou}}, \bibinfo {author}
  {\bibfnamefont {A.}~\bibnamefont {Brusch}}, \bibinfo {author} {\bibfnamefont
  {L.}~\bibnamefont {Lorini}}, \bibinfo {author} {\bibfnamefont {W.~H.}\
  \bibnamefont {Oskay}}, \bibinfo {author} {\bibfnamefont {R.~E.}\ \bibnamefont
  {Drullinger}}, \bibinfo {author} {\bibfnamefont {T.~M.}\ \bibnamefont
  {Fortier}}, \bibinfo {author} {\bibfnamefont {J.~E.}\ \bibnamefont
  {Stalnaker}}, \bibinfo {author} {\bibfnamefont {S.~A.}\ \bibnamefont
  {Diddams}}, \bibinfo {author} {\bibfnamefont {W.~C.}\ \bibnamefont {Swann}},
  \bibinfo {author} {\bibfnamefont {N.~R.}\ \bibnamefont {Newbury}}, \bibinfo
  {author} {\bibfnamefont {W.~M.}\ \bibnamefont {Itano}}, \bibinfo {author}
  {\bibfnamefont {D.~J.}\ \bibnamefont {Wineland}}, \ and\ \bibinfo {author}
  {\bibfnamefont {J.~C.}\ \bibnamefont {Bergquist}},\ }\href {\doibase
  10.1126/science.1154622} {\bibfield  {journal} {\bibinfo  {journal}
  {Science}\ }\textbf {\bibinfo {volume} {319}},\ \bibinfo {pages} {1808}
  (\bibinfo {year} {2008})}\BibitemShut {NoStop}%
\bibitem [{\citenamefont {Chou}\ \emph {et~al.}(2010)\citenamefont {Chou},
  \citenamefont {Hume}, \citenamefont {Koelemeij}, \citenamefont {Wineland},\
  and\ \citenamefont {Rosenband}}]{ChouPRL10}%
  \BibitemOpen
  \bibfield  {author} {\bibinfo {author} {\bibfnamefont {C.~W.}\ \bibnamefont
  {Chou}}, \bibinfo {author} {\bibfnamefont {D.~B.}\ \bibnamefont {Hume}},
  \bibinfo {author} {\bibfnamefont {J.~C.~J.}\ \bibnamefont {Koelemeij}},
  \bibinfo {author} {\bibfnamefont {D.~J.}\ \bibnamefont {Wineland}}, \ and\
  \bibinfo {author} {\bibfnamefont {T.}~\bibnamefont {Rosenband}},\ }\href
  {\doibase 10.1103/PhysRevLett.104.070802} {\bibfield  {journal} {\bibinfo
  {journal} {Phys. Rev. Lett.}\ }\textbf {\bibinfo {volume} {104}},\ \bibinfo
  {pages} {070802} (\bibinfo {year} {2010})}\BibitemShut {NoStop}%
\bibitem [{\citenamefont {Huntemann}\ \emph {et~al.}(2016)\citenamefont
  {Huntemann}, \citenamefont {Sanner}, \citenamefont {Lipphardt}, \citenamefont
  {Tamm},\ and\ \citenamefont {Peik}}]{HuntemannPRL16}%
  \BibitemOpen
  \bibfield  {author} {\bibinfo {author} {\bibfnamefont {N.}~\bibnamefont
  {Huntemann}}, \bibinfo {author} {\bibfnamefont {C.}~\bibnamefont {Sanner}},
  \bibinfo {author} {\bibfnamefont {B.}~\bibnamefont {Lipphardt}}, \bibinfo
  {author} {\bibfnamefont {C.}~\bibnamefont {Tamm}}, \ and\ \bibinfo {author}
  {\bibfnamefont {E.}~\bibnamefont {Peik}},\ }\href {\doibase
  10.1103/PhysRevLett.116.063001} {\bibfield  {journal} {\bibinfo  {journal}
  {Phys. Rev. Lett.}\ }\textbf {\bibinfo {volume} {116}},\ \bibinfo {pages}
  {063001} (\bibinfo {year} {2016})}\BibitemShut {NoStop}%
\bibitem [{\citenamefont {Chen}\ \emph {et~al.}(2017)\citenamefont {Chen},
  \citenamefont {Brewer}, \citenamefont {Chou}, \citenamefont {Wineland},
  \citenamefont {Leibrandt},\ and\ \citenamefont {Hume}}]{ChenPRL17}%
  \BibitemOpen
  \bibfield  {author} {\bibinfo {author} {\bibfnamefont {J.-S.}\ \bibnamefont
  {Chen}}, \bibinfo {author} {\bibfnamefont {S.~M.}\ \bibnamefont {Brewer}},
  \bibinfo {author} {\bibfnamefont {C.~W.}\ \bibnamefont {Chou}}, \bibinfo
  {author} {\bibfnamefont {D.~J.}\ \bibnamefont {Wineland}}, \bibinfo {author}
  {\bibfnamefont {D.~R.}\ \bibnamefont {Leibrandt}}, \ and\ \bibinfo {author}
  {\bibfnamefont {D.~B.}\ \bibnamefont {Hume}},\ }\href {\doibase
  10.1103/PhysRevLett.118.053002} {\bibfield  {journal} {\bibinfo  {journal}
  {Phys. Rev. Lett.}\ }\textbf {\bibinfo {volume} {118}},\ \bibinfo {pages}
  {053002} (\bibinfo {year} {2017})}\BibitemShut {NoStop}%
\bibitem [{\citenamefont {Brewer}\ \emph
  {et~al.}(2019{\natexlab{a}})\citenamefont {Brewer}, \citenamefont {Chen},
  \citenamefont {Hankin}, \citenamefont {Clements}, \citenamefont {Chou},
  \citenamefont {Wineland}, \citenamefont {Hume},\ and\ \citenamefont
  {Leibrandt}}]{Brewer2019a}%
  \BibitemOpen
  \bibfield  {author} {\bibinfo {author} {\bibfnamefont {S.~M.}\ \bibnamefont
  {Brewer}}, \bibinfo {author} {\bibfnamefont {J.-S.}\ \bibnamefont {Chen}},
  \bibinfo {author} {\bibfnamefont {A.~M.}\ \bibnamefont {Hankin}}, \bibinfo
  {author} {\bibfnamefont {E.~R.}\ \bibnamefont {Clements}}, \bibinfo {author}
  {\bibfnamefont {C.~W.}\ \bibnamefont {Chou}}, \bibinfo {author}
  {\bibfnamefont {D.~J.}\ \bibnamefont {Wineland}}, \bibinfo {author}
  {\bibfnamefont {D.~B.}\ \bibnamefont {Hume}}, \ and\ \bibinfo {author}
  {\bibfnamefont {D.~R.}\ \bibnamefont {Leibrandt}},\ }\href {\doibase
  10.1103/PhysRevLett.123.033201} {\bibfield  {journal} {\bibinfo  {journal}
  {Phys. Rev. Lett.}\ }\textbf {\bibinfo {volume} {123}},\ \bibinfo {pages}
  {033201} (\bibinfo {year} {2019}{\natexlab{a}})}\BibitemShut {NoStop}%
\bibitem [{\citenamefont {H\"arter}\ and\ \citenamefont
  {Denschlag}(2014)}]{HarterCP14}%
  \BibitemOpen
  \bibfield  {author} {\bibinfo {author} {\bibfnamefont {A.}~\bibnamefont
  {H\"arter}}\ and\ \bibinfo {author} {\bibfnamefont {J.~H.}\ \bibnamefont
  {Denschlag}},\ }\href {\doibase 10.1080/00107514.2013.854618} {\bibfield
  {journal} {\bibinfo  {journal} {Contemp. Phys.}\ }\textbf {\bibinfo {volume}
  {55}},\ \bibinfo {pages} {33} (\bibinfo {year} {2014})}\BibitemShut {NoStop}%
\bibitem [{\citenamefont {C{\^o}t{\'e}}(2016)}]{CoteAAMOP16}%
  \BibitemOpen
  \bibfield  {author} {\bibinfo {author} {\bibfnamefont {R.}~\bibnamefont
  {C{\^o}t{\'e}}},\ }\href {\doibase 10.1016/bs.aamop.2016.04.004} {\bibfield
  {journal} {\bibinfo  {journal} {Adv. At. Mol. Opt. Phys.}\ }\textbf {\bibinfo
  {volume} {65}},\ \bibinfo {pages} {67} (\bibinfo {year} {2016})}\BibitemShut
  {NoStop}%
\bibitem [{\citenamefont {Tomza}\ \emph {et~al.}(2019)\citenamefont {Tomza},
  \citenamefont {Jachymski}, \citenamefont {Gerritsma}, \citenamefont
  {Negretti}, \citenamefont {Calarco}, \citenamefont {Idziaszek},\ and\
  \citenamefont {Julienne}}]{TomzaRMP18}%
  \BibitemOpen
  \bibfield  {author} {\bibinfo {author} {\bibfnamefont {M.}~\bibnamefont
  {Tomza}}, \bibinfo {author} {\bibfnamefont {K.}~\bibnamefont {Jachymski}},
  \bibinfo {author} {\bibfnamefont {R.}~\bibnamefont {Gerritsma}}, \bibinfo
  {author} {\bibfnamefont {A.}~\bibnamefont {Negretti}}, \bibinfo {author}
  {\bibfnamefont {T.}~\bibnamefont {Calarco}}, \bibinfo {author} {\bibfnamefont
  {Z.}~\bibnamefont {Idziaszek}}, \ and\ \bibinfo {author} {\bibfnamefont
  {P.~S.}\ \bibnamefont {Julienne}},\ }\href {\doibase
  10.1103/RevModPhys.91.035001} {\bibfield  {journal} {\bibinfo  {journal}
  {Rev. Mod. Phys.}\ }\textbf {\bibinfo {volume} {91}},\ \bibinfo {pages}
  {035001} (\bibinfo {year} {2019})}\BibitemShut {NoStop}%
\bibitem [{\citenamefont {Bloch}\ \emph {et~al.}(2008)\citenamefont {Bloch},
  \citenamefont {Dalibard},\ and\ \citenamefont {Zwerger}}]{BlochRMP08}%
  \BibitemOpen
  \bibfield  {author} {\bibinfo {author} {\bibfnamefont {I.}~\bibnamefont
  {Bloch}}, \bibinfo {author} {\bibfnamefont {J.}~\bibnamefont {Dalibard}}, \
  and\ \bibinfo {author} {\bibfnamefont {W.}~\bibnamefont {Zwerger}},\ }\href
  {\doibase 10.1103/RevModPhys.80.885} {\bibfield  {journal} {\bibinfo
  {journal} {Rev. Mod. Phys.}\ }\textbf {\bibinfo {volume} {80}},\ \bibinfo
  {pages} {885} (\bibinfo {year} {2008})}\BibitemShut {NoStop}%
\bibitem [{\citenamefont {Gross}\ and\ \citenamefont
  {Bloch}(2017)}]{GrossScience17}%
  \BibitemOpen
  \bibfield  {author} {\bibinfo {author} {\bibfnamefont {C.}~\bibnamefont
  {Gross}}\ and\ \bibinfo {author} {\bibfnamefont {I.}~\bibnamefont {Bloch}},\
  }\href {\doibase 10.1126/science.aal3837} {\bibfield  {journal} {\bibinfo
  {journal} {Science}\ }\textbf {\bibinfo {volume} {357}},\ \bibinfo {pages}
  {995} (\bibinfo {year} {2017})}\BibitemShut {NoStop}%
\bibitem [{\citenamefont {C\^ot\'e}\ and\ \citenamefont
  {Dalgarno}(2000)}]{CotePRA00}%
  \BibitemOpen
  \bibfield  {author} {\bibinfo {author} {\bibfnamefont {R.}~\bibnamefont
  {C\^ot\'e}}\ and\ \bibinfo {author} {\bibfnamefont {A.}~\bibnamefont
  {Dalgarno}},\ }\href {\doibase 10.1103/PhysRevA.62.012709} {\bibfield
  {journal} {\bibinfo  {journal} {Phys. Rev. A}\ }\textbf {\bibinfo {volume}
  {62}},\ \bibinfo {pages} {012709} (\bibinfo {year} {2000})}\BibitemShut
  {NoStop}%
\bibitem [{\citenamefont {Zipkes}\ \emph {et~al.}(2010)\citenamefont {Zipkes},
  \citenamefont {Palzer}, \citenamefont {Sias},\ and\ \citenamefont
  {K{\"o}hl}}]{ZipkesNature10}%
  \BibitemOpen
  \bibfield  {author} {\bibinfo {author} {\bibfnamefont {C.}~\bibnamefont
  {Zipkes}}, \bibinfo {author} {\bibfnamefont {S.}~\bibnamefont {Palzer}},
  \bibinfo {author} {\bibfnamefont {C.}~\bibnamefont {Sias}}, \ and\ \bibinfo
  {author} {\bibfnamefont {M.}~\bibnamefont {K{\"o}hl}},\ }\href {\doibase
  10.1038/nature08865} {\bibfield  {journal} {\bibinfo  {journal} {Nature}\
  }\textbf {\bibinfo {volume} {464}},\ \bibinfo {pages} {388} (\bibinfo {year}
  {2010})}\BibitemShut {NoStop}%
\bibitem [{\citenamefont {Ravi}\ \emph {et~al.}(2012)\citenamefont {Ravi},
  \citenamefont {Lee}, \citenamefont {Sharma}, \citenamefont {Werth},\ and\
  \citenamefont {Rangwala}}]{RaviNatCommun12}%
  \BibitemOpen
  \bibfield  {author} {\bibinfo {author} {\bibfnamefont {K.}~\bibnamefont
  {Ravi}}, \bibinfo {author} {\bibfnamefont {S.}~\bibnamefont {Lee}}, \bibinfo
  {author} {\bibfnamefont {A.}~\bibnamefont {Sharma}}, \bibinfo {author}
  {\bibfnamefont {G.}~\bibnamefont {Werth}}, \ and\ \bibinfo {author}
  {\bibfnamefont {S.}~\bibnamefont {Rangwala}},\ }\href {\doibase
  10.1038/ncomms2131} {\bibfield  {journal} {\bibinfo  {journal} {Nat.
  Commun.}\ }\textbf {\bibinfo {volume} {3}},\ \bibinfo {pages} {1126}
  (\bibinfo {year} {2012})}\BibitemShut {NoStop}%
\bibitem [{\citenamefont {H\"oltkemeier}\ \emph {et~al.}(2016)\citenamefont
  {H\"oltkemeier}, \citenamefont {Weckesser}, \citenamefont {L\'opez-Carrera},\
  and\ \citenamefont {Weidem\"uller}}]{HoltkemeierPRL16}%
  \BibitemOpen
  \bibfield  {author} {\bibinfo {author} {\bibfnamefont {B.}~\bibnamefont
  {H\"oltkemeier}}, \bibinfo {author} {\bibfnamefont {P.}~\bibnamefont
  {Weckesser}}, \bibinfo {author} {\bibfnamefont {H.}~\bibnamefont
  {L\'opez-Carrera}}, \ and\ \bibinfo {author} {\bibfnamefont {M.}~\bibnamefont
  {Weidem\"uller}},\ }\href {\doibase 10.1103/PhysRevLett.116.233003}
  {\bibfield  {journal} {\bibinfo  {journal} {Phys. Rev. Lett.}\ }\textbf
  {\bibinfo {volume} {116}},\ \bibinfo {pages} {233003} (\bibinfo {year}
  {2016})}\BibitemShut {NoStop}%
\bibitem [{\citenamefont {Petrov}\ \emph {et~al.}(2017)\citenamefont {Petrov},
  \citenamefont {Makrides},\ and\ \citenamefont {Kotochigova}}]{PetrovJCP17}%
  \BibitemOpen
  \bibfield  {author} {\bibinfo {author} {\bibfnamefont {A.}~\bibnamefont
  {Petrov}}, \bibinfo {author} {\bibfnamefont {C.}~\bibnamefont {Makrides}}, \
  and\ \bibinfo {author} {\bibfnamefont {S.}~\bibnamefont {Kotochigova}},\
  }\href {\doibase 10.1063/1.4976972} {\bibfield  {journal} {\bibinfo
  {journal} {J. Chem. Phys.}\ }\textbf {\bibinfo {volume} {146}},\ \bibinfo
  {pages} {084304} (\bibinfo {year} {2017})}\BibitemShut {NoStop}%
\bibitem [{\citenamefont {Feldker}\ \emph {et~al.}(2019)\citenamefont
  {Feldker}, \citenamefont {F{\"u}rst}, \citenamefont {Hirzler}, \citenamefont
  {Ewald}, \citenamefont {Mazzanti}, \citenamefont {Wiater}, \citenamefont
  {Tomza},\ and\ \citenamefont {Gerritsma}}]{Feldker2019}%
  \BibitemOpen
  \bibfield  {author} {\bibinfo {author} {\bibfnamefont {T.}~\bibnamefont
  {Feldker}}, \bibinfo {author} {\bibfnamefont {H.}~\bibnamefont {F{\"u}rst}},
  \bibinfo {author} {\bibfnamefont {H.}~\bibnamefont {Hirzler}}, \bibinfo
  {author} {\bibfnamefont {N.}~\bibnamefont {Ewald}}, \bibinfo {author}
  {\bibfnamefont {M.}~\bibnamefont {Mazzanti}}, \bibinfo {author}
  {\bibfnamefont {D.}~\bibnamefont {Wiater}}, \bibinfo {author} {\bibfnamefont
  {M.}~\bibnamefont {Tomza}}, \ and\ \bibinfo {author} {\bibfnamefont
  {R.}~\bibnamefont {Gerritsma}},\ }\href@noop {} {\bibfield  {journal}
  {\bibinfo  {journal} {arXiv preprint arXiv:1907.10926}\ } (\bibinfo {year}
  {2019})}\BibitemShut {NoStop}%
\bibitem [{\citenamefont {Hall}\ and\ \citenamefont
  {Willitsch}(2012)}]{HallPRL12}%
  \BibitemOpen
  \bibfield  {author} {\bibinfo {author} {\bibfnamefont {F.~H.~J.}\
  \bibnamefont {Hall}}\ and\ \bibinfo {author} {\bibfnamefont {S.}~\bibnamefont
  {Willitsch}},\ }\href {\doibase 10.1103/PhysRevLett.109.233202} {\bibfield
  {journal} {\bibinfo  {journal} {Phys. Rev. Lett.}\ }\textbf {\bibinfo
  {volume} {109}},\ \bibinfo {pages} {233202} (\bibinfo {year}
  {2012})}\BibitemShut {NoStop}%
\bibitem [{\citenamefont {Makarov}\ \emph {et~al.}(2003)\citenamefont
  {Makarov}, \citenamefont {C\^ot\'e}, \citenamefont {Michels},\ and\
  \citenamefont {Smith}}]{MakarovPRA03}%
  \BibitemOpen
  \bibfield  {author} {\bibinfo {author} {\bibfnamefont {O.~P.}\ \bibnamefont
  {Makarov}}, \bibinfo {author} {\bibfnamefont {R.}~\bibnamefont {C\^ot\'e}},
  \bibinfo {author} {\bibfnamefont {H.}~\bibnamefont {Michels}}, \ and\
  \bibinfo {author} {\bibfnamefont {W.~W.}\ \bibnamefont {Smith}},\ }\href
  {\doibase 10.1103/PhysRevA.67.042705} {\bibfield  {journal} {\bibinfo
  {journal} {Phys. Rev. A}\ }\textbf {\bibinfo {volume} {67}},\ \bibinfo
  {pages} {042705} (\bibinfo {year} {2003})}\BibitemShut {NoStop}%
\bibitem [{\citenamefont {Ratschbacher}\ \emph {et~al.}(2012)\citenamefont
  {Ratschbacher}, \citenamefont {Zipkes}, \citenamefont {Sias},\ and\
  \citenamefont {Kohl}}]{RatschbacherNatPhys12}%
  \BibitemOpen
  \bibfield  {author} {\bibinfo {author} {\bibfnamefont {L.}~\bibnamefont
  {Ratschbacher}}, \bibinfo {author} {\bibfnamefont {C.}~\bibnamefont
  {Zipkes}}, \bibinfo {author} {\bibfnamefont {C.}~\bibnamefont {Sias}}, \ and\
  \bibinfo {author} {\bibfnamefont {M.}~\bibnamefont {Kohl}},\ }\href {\doibase
  10.1038/nphys2373} {\bibfield  {journal} {\bibinfo  {journal} {Nat. Phys.}\
  }\textbf {\bibinfo {volume} {8}},\ \bibinfo {pages} {649} (\bibinfo {year}
  {2012})}\BibitemShut {NoStop}%
\bibitem [{\citenamefont {F\"urst}\ \emph {et~al.}(2018)\citenamefont
  {F\"urst}, \citenamefont {Feldker}, \citenamefont {Ewald}, \citenamefont
  {Joger}, \citenamefont {Tomza},\ and\ \citenamefont
  {Gerritsma}}]{FurstPRA18}%
  \BibitemOpen
  \bibfield  {author} {\bibinfo {author} {\bibfnamefont {H.}~\bibnamefont
  {F\"urst}}, \bibinfo {author} {\bibfnamefont {T.}~\bibnamefont {Feldker}},
  \bibinfo {author} {\bibfnamefont {N.~V.}\ \bibnamefont {Ewald}}, \bibinfo
  {author} {\bibfnamefont {J.}~\bibnamefont {Joger}}, \bibinfo {author}
  {\bibfnamefont {M.}~\bibnamefont {Tomza}}, \ and\ \bibinfo {author}
  {\bibfnamefont {R.}~\bibnamefont {Gerritsma}},\ }\href {\doibase
  10.1103/PhysRevA.98.012713} {\bibfield  {journal} {\bibinfo  {journal} {Phys.
  Rev. A}\ }\textbf {\bibinfo {volume} {98}},\ \bibinfo {pages} {012713}
  (\bibinfo {year} {2018})}\BibitemShut {NoStop}%
\bibitem [{\citenamefont {Sikorsky}\ \emph
  {et~al.}(2018{\natexlab{a}})\citenamefont {Sikorsky}, \citenamefont {Morita},
  \citenamefont {Meir}, \citenamefont {Buchachenko}, \citenamefont
  {Ben-shlomi}, \citenamefont {Akerman}, \citenamefont {Narevicius},
  \citenamefont {Tscherbul},\ and\ \citenamefont {Ozeri}}]{SikorskyPRL18}%
  \BibitemOpen
  \bibfield  {author} {\bibinfo {author} {\bibfnamefont {T.}~\bibnamefont
  {Sikorsky}}, \bibinfo {author} {\bibfnamefont {M.}~\bibnamefont {Morita}},
  \bibinfo {author} {\bibfnamefont {Z.}~\bibnamefont {Meir}}, \bibinfo {author}
  {\bibfnamefont {A.~A.}\ \bibnamefont {Buchachenko}}, \bibinfo {author}
  {\bibfnamefont {R.}~\bibnamefont {Ben-shlomi}}, \bibinfo {author}
  {\bibfnamefont {N.}~\bibnamefont {Akerman}}, \bibinfo {author} {\bibfnamefont
  {E.}~\bibnamefont {Narevicius}}, \bibinfo {author} {\bibfnamefont {T.~V.}\
  \bibnamefont {Tscherbul}}, \ and\ \bibinfo {author} {\bibfnamefont
  {R.}~\bibnamefont {Ozeri}},\ }\href {\doibase 10.1103/PhysRevLett.121.173402}
  {\bibfield  {journal} {\bibinfo  {journal} {Phys. Rev. Lett.}\ }\textbf
  {\bibinfo {volume} {121}},\ \bibinfo {pages} {173402} (\bibinfo {year}
  {2018}{\natexlab{a}})}\BibitemShut {NoStop}%
\bibitem [{\citenamefont {C\^ot\'e}\ and\ \citenamefont
  {Simbotin}(2018)}]{CotePRL18}%
  \BibitemOpen
  \bibfield  {author} {\bibinfo {author} {\bibfnamefont {R.}~\bibnamefont
  {C\^ot\'e}}\ and\ \bibinfo {author} {\bibfnamefont {I.}~\bibnamefont
  {Simbotin}},\ }\href {\doibase 10.1103/PhysRevLett.121.173401} {\bibfield
  {journal} {\bibinfo  {journal} {Phys. Rev. Lett.}\ }\textbf {\bibinfo
  {volume} {121}},\ \bibinfo {pages} {173401} (\bibinfo {year}
  {2018})}\BibitemShut {NoStop}%
\bibitem [{\citenamefont {Saito}\ \emph {et~al.}(2017)\citenamefont {Saito},
  \citenamefont {Haze}, \citenamefont {Sasakawa}, \citenamefont {Nakai},
  \citenamefont {Raoult}, \citenamefont {Da~Silva}, \citenamefont {Dulieu},\
  and\ \citenamefont {Mukaiyama}}]{SaitoPRA17}%
  \BibitemOpen
  \bibfield  {author} {\bibinfo {author} {\bibfnamefont {R.}~\bibnamefont
  {Saito}}, \bibinfo {author} {\bibfnamefont {S.}~\bibnamefont {Haze}},
  \bibinfo {author} {\bibfnamefont {M.}~\bibnamefont {Sasakawa}}, \bibinfo
  {author} {\bibfnamefont {R.}~\bibnamefont {Nakai}}, \bibinfo {author}
  {\bibfnamefont {M.}~\bibnamefont {Raoult}}, \bibinfo {author} {\bibfnamefont
  {H.}~\bibnamefont {Da~Silva}}, \bibinfo {author} {\bibfnamefont
  {O.}~\bibnamefont {Dulieu}}, \ and\ \bibinfo {author} {\bibfnamefont
  {T.}~\bibnamefont {Mukaiyama}},\ }\href {\doibase 10.1103/PhysRevA.95.032709}
  {\bibfield  {journal} {\bibinfo  {journal} {Phys. Rev. A}\ }\textbf {\bibinfo
  {volume} {95}},\ \bibinfo {pages} {032709} (\bibinfo {year}
  {2017})}\BibitemShut {NoStop}%
\bibitem [{\citenamefont {Sikorsky}\ \emph
  {et~al.}(2018{\natexlab{b}})\citenamefont {Sikorsky}, \citenamefont {Meir},
  \citenamefont {Ben-shlomi}, \citenamefont {Akerman},\ and\ \citenamefont
  {Ozeri}}]{SikorskyNC18}%
  \BibitemOpen
  \bibfield  {author} {\bibinfo {author} {\bibfnamefont {T.}~\bibnamefont
  {Sikorsky}}, \bibinfo {author} {\bibfnamefont {Z.}~\bibnamefont {Meir}},
  \bibinfo {author} {\bibfnamefont {R.}~\bibnamefont {Ben-shlomi}}, \bibinfo
  {author} {\bibfnamefont {N.}~\bibnamefont {Akerman}}, \ and\ \bibinfo
  {author} {\bibfnamefont {R.}~\bibnamefont {Ozeri}},\ }\href {\doibase
  10.1038/s41467-018-03373-y} {\bibfield  {journal} {\bibinfo  {journal} {Nat.
  Comm.}\ }\textbf {\bibinfo {volume} {9}},\ \bibinfo {pages} {920} (\bibinfo
  {year} {2018}{\natexlab{b}})}\BibitemShut {NoStop}%
\bibitem [{\citenamefont {Bissbort}\ \emph {et~al.}(2013)\citenamefont
  {Bissbort}, \citenamefont {Cocks}, \citenamefont {Negretti}, \citenamefont
  {Idziaszek}, \citenamefont {Calarco}, \citenamefont {Schmidt-Kaler},
  \citenamefont {Hofstetter},\ and\ \citenamefont {Gerritsma}}]{BissbortPRL13}%
  \BibitemOpen
  \bibfield  {author} {\bibinfo {author} {\bibfnamefont {U.}~\bibnamefont
  {Bissbort}}, \bibinfo {author} {\bibfnamefont {D.}~\bibnamefont {Cocks}},
  \bibinfo {author} {\bibfnamefont {A.}~\bibnamefont {Negretti}}, \bibinfo
  {author} {\bibfnamefont {Z.}~\bibnamefont {Idziaszek}}, \bibinfo {author}
  {\bibfnamefont {T.}~\bibnamefont {Calarco}}, \bibinfo {author} {\bibfnamefont
  {F.}~\bibnamefont {Schmidt-Kaler}}, \bibinfo {author} {\bibfnamefont
  {W.}~\bibnamefont {Hofstetter}}, \ and\ \bibinfo {author} {\bibfnamefont
  {R.}~\bibnamefont {Gerritsma}},\ }\href {\doibase
  10.1103/PhysRevLett.111.080501} {\bibfield  {journal} {\bibinfo  {journal}
  {Phys. Rev. Lett.}\ }\textbf {\bibinfo {volume} {111}},\ \bibinfo {pages}
  {080501} (\bibinfo {year} {2013})}\BibitemShut {NoStop}%
\bibitem [{\citenamefont {Doerk}\ \emph {et~al.}(2010)\citenamefont {Doerk},
  \citenamefont {Idziaszek},\ and\ \citenamefont {Calarco}}]{DoerkPRA10}%
  \BibitemOpen
  \bibfield  {author} {\bibinfo {author} {\bibfnamefont {H.}~\bibnamefont
  {Doerk}}, \bibinfo {author} {\bibfnamefont {Z.}~\bibnamefont {Idziaszek}}, \
  and\ \bibinfo {author} {\bibfnamefont {T.}~\bibnamefont {Calarco}},\ }\href
  {\doibase 10.1103/PhysRevA.81.012708} {\bibfield  {journal} {\bibinfo
  {journal} {Phys. Rev. A}\ }\textbf {\bibinfo {volume} {81}},\ \bibinfo
  {pages} {012708} (\bibinfo {year} {2010})}\BibitemShut {NoStop}%
\bibitem [{\citenamefont {Grier}\ \emph {et~al.}(2009)\citenamefont {Grier},
  \citenamefont {Cetina}, \citenamefont {Oru\ifmmode \check{c}\else
  \v{c}\fi{}evi\ifmmode~\acute{c}\else \'{c}\fi{}},\ and\ \citenamefont
  {Vuleti\ifmmode~\acute{c}\else \'{c}\fi{}}}]{GrierPRL09}%
  \BibitemOpen
  \bibfield  {author} {\bibinfo {author} {\bibfnamefont {A.~T.}\ \bibnamefont
  {Grier}}, \bibinfo {author} {\bibfnamefont {M.}~\bibnamefont {Cetina}},
  \bibinfo {author} {\bibfnamefont {F.}~\bibnamefont {Oru\ifmmode
  \check{c}\else \v{c}\fi{}evi\ifmmode~\acute{c}\else \'{c}\fi{}}}, \ and\
  \bibinfo {author} {\bibfnamefont {V.}~\bibnamefont
  {Vuleti\ifmmode~\acute{c}\else \'{c}\fi{}}},\ }\href {\doibase
  10.1103/PhysRevLett.102.223201} {\bibfield  {journal} {\bibinfo  {journal}
  {Phys. Rev. Lett.}\ }\textbf {\bibinfo {volume} {102}},\ \bibinfo {pages}
  {223201} (\bibinfo {year} {2009})}\BibitemShut {NoStop}%
\bibitem [{\citenamefont {Hall}\ \emph {et~al.}(2013)\citenamefont {Hall},
  \citenamefont {Eberle}, \citenamefont {Hegi}, \citenamefont {Raoult},
  \citenamefont {Aymar}, \citenamefont {Dulieu},\ and\ \citenamefont
  {Willitsch}}]{HallMP13a}%
  \BibitemOpen
  \bibfield  {author} {\bibinfo {author} {\bibfnamefont {F.~H.}\ \bibnamefont
  {Hall}}, \bibinfo {author} {\bibfnamefont {P.}~\bibnamefont {Eberle}},
  \bibinfo {author} {\bibfnamefont {G.}~\bibnamefont {Hegi}}, \bibinfo {author}
  {\bibfnamefont {M.}~\bibnamefont {Raoult}}, \bibinfo {author} {\bibfnamefont
  {M.}~\bibnamefont {Aymar}}, \bibinfo {author} {\bibfnamefont
  {O.}~\bibnamefont {Dulieu}}, \ and\ \bibinfo {author} {\bibfnamefont
  {S.}~\bibnamefont {Willitsch}},\ }\href {\doibase
  10.1080/00268976.2013.780107} {\bibfield  {journal} {\bibinfo  {journal}
  {Mol. Phys.}\ }\textbf {\bibinfo {volume} {111}},\ \bibinfo {pages} {2020}
  (\bibinfo {year} {2013})}\BibitemShut {NoStop}%
\bibitem [{\citenamefont {Sullivan}\ \emph {et~al.}(2012)\citenamefont
  {Sullivan}, \citenamefont {Rellergert}, \citenamefont {Kotochigova},\ and\
  \citenamefont {Hudson}}]{SullivanPRL12}%
  \BibitemOpen
  \bibfield  {author} {\bibinfo {author} {\bibfnamefont {S.~T.}\ \bibnamefont
  {Sullivan}}, \bibinfo {author} {\bibfnamefont {W.~G.}\ \bibnamefont
  {Rellergert}}, \bibinfo {author} {\bibfnamefont {S.}~\bibnamefont
  {Kotochigova}}, \ and\ \bibinfo {author} {\bibfnamefont {E.~R.}\ \bibnamefont
  {Hudson}},\ }\href {\doibase 10.1103/PhysRevLett.109.223002} {\bibfield
  {journal} {\bibinfo  {journal} {Phys. Rev. Lett.}\ }\textbf {\bibinfo
  {volume} {109}},\ \bibinfo {pages} {223002} (\bibinfo {year}
  {2012})}\BibitemShut {NoStop}%
\bibitem [{\citenamefont {Rellergert}\ \emph {et~al.}(2011)\citenamefont
  {Rellergert}, \citenamefont {Sullivan}, \citenamefont {Kotochigova},
  \citenamefont {Petrov}, \citenamefont {Chen}, \citenamefont {Schowalter},\
  and\ \citenamefont {Hudson}}]{RellergertPRL11}%
  \BibitemOpen
  \bibfield  {author} {\bibinfo {author} {\bibfnamefont {W.~G.}\ \bibnamefont
  {Rellergert}}, \bibinfo {author} {\bibfnamefont {S.~T.}\ \bibnamefont
  {Sullivan}}, \bibinfo {author} {\bibfnamefont {S.}~\bibnamefont
  {Kotochigova}}, \bibinfo {author} {\bibfnamefont {A.}~\bibnamefont {Petrov}},
  \bibinfo {author} {\bibfnamefont {K.}~\bibnamefont {Chen}}, \bibinfo {author}
  {\bibfnamefont {S.~J.}\ \bibnamefont {Schowalter}}, \ and\ \bibinfo {author}
  {\bibfnamefont {E.~R.}\ \bibnamefont {Hudson}},\ }\href {\doibase
  10.1103/PhysRevLett.107.243201} {\bibfield  {journal} {\bibinfo  {journal}
  {Phys. Rev. Lett.}\ }\textbf {\bibinfo {volume} {107}},\ \bibinfo {pages}
  {243201} (\bibinfo {year} {2011})}\BibitemShut {NoStop}%
\bibitem [{\citenamefont {Haze}\ \emph {et~al.}(2013)\citenamefont {Haze},
  \citenamefont {Hata}, \citenamefont {Fujinaga},\ and\ \citenamefont
  {Mukaiyama}}]{HazePRA13}%
  \BibitemOpen
  \bibfield  {author} {\bibinfo {author} {\bibfnamefont {S.}~\bibnamefont
  {Haze}}, \bibinfo {author} {\bibfnamefont {S.}~\bibnamefont {Hata}}, \bibinfo
  {author} {\bibfnamefont {M.}~\bibnamefont {Fujinaga}}, \ and\ \bibinfo
  {author} {\bibfnamefont {T.}~\bibnamefont {Mukaiyama}},\ }\href {\doibase
  10.1103/PhysRevA.87.052715} {\bibfield  {journal} {\bibinfo  {journal} {Phys.
  Rev. A}\ }\textbf {\bibinfo {volume} {87}},\ \bibinfo {pages} {052715}
  (\bibinfo {year} {2013})}\BibitemShut {NoStop}%
\bibitem [{\citenamefont {Hall}\ \emph {et~al.}(2011)\citenamefont {Hall},
  \citenamefont {Aymar}, \citenamefont {Bouloufa-Maafa}, \citenamefont
  {Dulieu},\ and\ \citenamefont {Willitsch}}]{HallPRL11}%
  \BibitemOpen
  \bibfield  {author} {\bibinfo {author} {\bibfnamefont {F.~H.~J.}\
  \bibnamefont {Hall}}, \bibinfo {author} {\bibfnamefont {M.}~\bibnamefont
  {Aymar}}, \bibinfo {author} {\bibfnamefont {N.}~\bibnamefont
  {Bouloufa-Maafa}}, \bibinfo {author} {\bibfnamefont {O.}~\bibnamefont
  {Dulieu}}, \ and\ \bibinfo {author} {\bibfnamefont {S.}~\bibnamefont
  {Willitsch}},\ }\href {\doibase 10.1103/PhysRevLett.107.243202} {\bibfield
  {journal} {\bibinfo  {journal} {Phys. Rev. Lett.}\ }\textbf {\bibinfo
  {volume} {107}},\ \bibinfo {pages} {243202} (\bibinfo {year}
  {2011})}\BibitemShut {NoStop}%
\bibitem [{\citenamefont {Smith}\ \emph {et~al.}(2014)\citenamefont {Smith},
  \citenamefont {Goodman}, \citenamefont {Sivarajah}, \citenamefont {Wells},
  \citenamefont {Banerjee}, \citenamefont {C\^ot\'e}, \citenamefont {Michels},
  \citenamefont {Mongtomery},\ and\ \citenamefont {Narducci}}]{SmithAPB14}%
  \BibitemOpen
  \bibfield  {author} {\bibinfo {author} {\bibfnamefont {W.}~\bibnamefont
  {Smith}}, \bibinfo {author} {\bibfnamefont {D.}~\bibnamefont {Goodman}},
  \bibinfo {author} {\bibfnamefont {I.}~\bibnamefont {Sivarajah}}, \bibinfo
  {author} {\bibfnamefont {J.}~\bibnamefont {Wells}}, \bibinfo {author}
  {\bibfnamefont {S.}~\bibnamefont {Banerjee}}, \bibinfo {author}
  {\bibfnamefont {R.}~\bibnamefont {C\^ot\'e}}, \bibinfo {author}
  {\bibfnamefont {H.}~\bibnamefont {Michels}}, \bibinfo {author} {\bibfnamefont
  {J.~A.}\ \bibnamefont {Mongtomery}}, \ and\ \bibinfo {author} {\bibfnamefont
  {F.}~\bibnamefont {Narducci}},\ }\href {\doibase 10.1007/s00340-013-5672-2}
  {\bibfield  {journal} {\bibinfo  {journal} {Appl. Phys. B}\ }\textbf
  {\bibinfo {volume} {114}},\ \bibinfo {pages} {75} (\bibinfo {year}
  {2014})}\BibitemShut {NoStop}%
\bibitem [{\citenamefont {Joger}\ \emph {et~al.}(2017)\citenamefont {Joger},
  \citenamefont {F\"urst}, \citenamefont {Ewald}, \citenamefont {Feldker},
  \citenamefont {Tomza},\ and\ \citenamefont {Gerritsma}}]{JogerPRA17}%
  \BibitemOpen
  \bibfield  {author} {\bibinfo {author} {\bibfnamefont {J.}~\bibnamefont
  {Joger}}, \bibinfo {author} {\bibfnamefont {H.}~\bibnamefont {F\"urst}},
  \bibinfo {author} {\bibfnamefont {N.}~\bibnamefont {Ewald}}, \bibinfo
  {author} {\bibfnamefont {T.}~\bibnamefont {Feldker}}, \bibinfo {author}
  {\bibfnamefont {M.}~\bibnamefont {Tomza}}, \ and\ \bibinfo {author}
  {\bibfnamefont {R.}~\bibnamefont {Gerritsma}},\ }\href {\doibase
  10.1103/PhysRevA.96.030703} {\bibfield  {journal} {\bibinfo  {journal} {Phys.
  Rev. A}\ }\textbf {\bibinfo {volume} {96}},\ \bibinfo {pages} {030703}
  (\bibinfo {year} {2017})}\BibitemShut {NoStop}%
\bibitem [{\citenamefont {Meir}\ \emph {et~al.}(2016)\citenamefont {Meir},
  \citenamefont {Sikorsky}, \citenamefont {Ben-shlomi}, \citenamefont
  {Akerman}, \citenamefont {Dallal},\ and\ \citenamefont {Ozeri}}]{MeirPRL16}%
  \BibitemOpen
  \bibfield  {author} {\bibinfo {author} {\bibfnamefont {Z.}~\bibnamefont
  {Meir}}, \bibinfo {author} {\bibfnamefont {T.}~\bibnamefont {Sikorsky}},
  \bibinfo {author} {\bibfnamefont {R.}~\bibnamefont {Ben-shlomi}}, \bibinfo
  {author} {\bibfnamefont {N.}~\bibnamefont {Akerman}}, \bibinfo {author}
  {\bibfnamefont {Y.}~\bibnamefont {Dallal}}, \ and\ \bibinfo {author}
  {\bibfnamefont {R.}~\bibnamefont {Ozeri}},\ }\href {\doibase
  10.1103/PhysRevLett.117.243401} {\bibfield  {journal} {\bibinfo  {journal}
  {Phys. Rev. Lett.}\ }\textbf {\bibinfo {volume} {117}},\ \bibinfo {pages}
  {243401} (\bibinfo {year} {2016})}\BibitemShut {NoStop}%
\bibitem [{\citenamefont {H\"arter}\ \emph {et~al.}(2012)\citenamefont
  {H\"arter}, \citenamefont {Kr\"ukow}, \citenamefont {Brunner}, \citenamefont
  {Schnitzler}, \citenamefont {Schmid},\ and\ \citenamefont
  {Hecker~Denschlag}}]{HartePRL12}%
  \BibitemOpen
  \bibfield  {author} {\bibinfo {author} {\bibfnamefont {A.}~\bibnamefont
  {H\"arter}}, \bibinfo {author} {\bibfnamefont {A.}~\bibnamefont {Kr\"ukow}},
  \bibinfo {author} {\bibfnamefont {A.}~\bibnamefont {Brunner}}, \bibinfo
  {author} {\bibfnamefont {W.}~\bibnamefont {Schnitzler}}, \bibinfo {author}
  {\bibfnamefont {S.}~\bibnamefont {Schmid}}, \ and\ \bibinfo {author}
  {\bibfnamefont {J.}~\bibnamefont {Hecker~Denschlag}},\ }\href {\doibase
  10.1103/PhysRevLett.109.123201} {\bibfield  {journal} {\bibinfo  {journal}
  {Phys. Rev. Lett.}\ }\textbf {\bibinfo {volume} {109}},\ \bibinfo {pages}
  {123201} (\bibinfo {year} {2012})}\BibitemShut {NoStop}%
\bibitem [{\citenamefont {Deiglmayr}\ \emph {et~al.}(2012)\citenamefont
  {Deiglmayr}, \citenamefont {G\"oritz}, \citenamefont {Best}, \citenamefont
  {Weidem\"uller},\ and\ \citenamefont {Wester}}]{DeiglmayrPRA12}%
  \BibitemOpen
  \bibfield  {author} {\bibinfo {author} {\bibfnamefont {J.}~\bibnamefont
  {Deiglmayr}}, \bibinfo {author} {\bibfnamefont {A.}~\bibnamefont {G\"oritz}},
  \bibinfo {author} {\bibfnamefont {T.}~\bibnamefont {Best}}, \bibinfo {author}
  {\bibfnamefont {M.}~\bibnamefont {Weidem\"uller}}, \ and\ \bibinfo {author}
  {\bibfnamefont {R.}~\bibnamefont {Wester}},\ }\href {\doibase
  10.1103/PhysRevA.86.043438} {\bibfield  {journal} {\bibinfo  {journal} {Phys.
  Rev. A}\ }\textbf {\bibinfo {volume} {86}},\ \bibinfo {pages} {043438}
  (\bibinfo {year} {2012})}\BibitemShut {NoStop}%
\bibitem [{\citenamefont {Jyothi}\ \emph {et~al.}(2016)\citenamefont {Jyothi},
  \citenamefont {Ray}, \citenamefont {Dutta}, \citenamefont {Allouche},
  \citenamefont {Vexiau}, \citenamefont {Dulieu},\ and\ \citenamefont
  {Rangwala}}]{JyothiPRL16}%
  \BibitemOpen
  \bibfield  {author} {\bibinfo {author} {\bibfnamefont {S.}~\bibnamefont
  {Jyothi}}, \bibinfo {author} {\bibfnamefont {T.}~\bibnamefont {Ray}},
  \bibinfo {author} {\bibfnamefont {S.}~\bibnamefont {Dutta}}, \bibinfo
  {author} {\bibfnamefont {A.~R.}\ \bibnamefont {Allouche}}, \bibinfo {author}
  {\bibfnamefont {R.}~\bibnamefont {Vexiau}}, \bibinfo {author} {\bibfnamefont
  {O.}~\bibnamefont {Dulieu}}, \ and\ \bibinfo {author} {\bibfnamefont {S.~A.}\
  \bibnamefont {Rangwala}},\ }\href {\doibase 10.1103/PhysRevLett.117.213002}
  {\bibfield  {journal} {\bibinfo  {journal} {Phys. Rev. Lett.}\ }\textbf
  {\bibinfo {volume} {117}},\ \bibinfo {pages} {213002} (\bibinfo {year}
  {2016})}\BibitemShut {NoStop}%
\bibitem [{\citenamefont {Kleinbach}\ \emph {et~al.}(2018)\citenamefont
  {Kleinbach}, \citenamefont {Engel}, \citenamefont {Dieterle}, \citenamefont
  {L\"ow}, \citenamefont {Pfau},\ and\ \citenamefont
  {Meinert}}]{KleinbachPRL18}%
  \BibitemOpen
  \bibfield  {author} {\bibinfo {author} {\bibfnamefont {K.~S.}\ \bibnamefont
  {Kleinbach}}, \bibinfo {author} {\bibfnamefont {F.}~\bibnamefont {Engel}},
  \bibinfo {author} {\bibfnamefont {T.}~\bibnamefont {Dieterle}}, \bibinfo
  {author} {\bibfnamefont {R.}~\bibnamefont {L\"ow}}, \bibinfo {author}
  {\bibfnamefont {T.}~\bibnamefont {Pfau}}, \ and\ \bibinfo {author}
  {\bibfnamefont {F.}~\bibnamefont {Meinert}},\ }\href {\doibase
  10.1103/PhysRevLett.120.193401} {\bibfield  {journal} {\bibinfo  {journal}
  {Phys. Rev. Lett.}\ }\textbf {\bibinfo {volume} {120}},\ \bibinfo {pages}
  {193401} (\bibinfo {year} {2018})}\BibitemShut {NoStop}%
\bibitem [{\citenamefont {Schmid}\ \emph {et~al.}(2018)\citenamefont {Schmid},
  \citenamefont {Veit}, \citenamefont {Zuber}, \citenamefont {L\"ow},
  \citenamefont {Pfau}, \citenamefont {Tarana},\ and\ \citenamefont
  {Tomza}}]{SchmidPRL18}%
  \BibitemOpen
  \bibfield  {author} {\bibinfo {author} {\bibfnamefont {T.}~\bibnamefont
  {Schmid}}, \bibinfo {author} {\bibfnamefont {C.}~\bibnamefont {Veit}},
  \bibinfo {author} {\bibfnamefont {N.}~\bibnamefont {Zuber}}, \bibinfo
  {author} {\bibfnamefont {R.}~\bibnamefont {L\"ow}}, \bibinfo {author}
  {\bibfnamefont {T.}~\bibnamefont {Pfau}}, \bibinfo {author} {\bibfnamefont
  {M.}~\bibnamefont {Tarana}}, \ and\ \bibinfo {author} {\bibfnamefont
  {M.}~\bibnamefont {Tomza}},\ }\href {\doibase 10.1103/PhysRevLett.120.153401}
  {\bibfield  {journal} {\bibinfo  {journal} {Phys. Rev. Lett.}\ }\textbf
  {\bibinfo {volume} {120}},\ \bibinfo {pages} {153401} (\bibinfo {year}
  {2018})}\BibitemShut {NoStop}%
\bibitem [{\citenamefont {Tomza}(2015)}]{TomzaPRA15b}%
  \BibitemOpen
  \bibfield  {author} {\bibinfo {author} {\bibfnamefont {M.}~\bibnamefont
  {Tomza}},\ }\href {\doibase 10.1103/PhysRevA.92.062701} {\bibfield  {journal}
  {\bibinfo  {journal} {Phys. Rev. A}\ }\textbf {\bibinfo {volume} {92}},\
  \bibinfo {pages} {062701} (\bibinfo {year} {2015})}\BibitemShut {NoStop}%
\bibitem [{\citenamefont {Tomza}\ \emph {et~al.}(2011)\citenamefont {Tomza},
  \citenamefont {Pawlowski}, \citenamefont {Jeziorska}, \citenamefont {Koch},\
  and\ \citenamefont {Moszynski}}]{TomzaPCCP11}%
  \BibitemOpen
  \bibfield  {author} {\bibinfo {author} {\bibfnamefont {M.}~\bibnamefont
  {Tomza}}, \bibinfo {author} {\bibfnamefont {F.}~\bibnamefont {Pawlowski}},
  \bibinfo {author} {\bibfnamefont {M.}~\bibnamefont {Jeziorska}}, \bibinfo
  {author} {\bibfnamefont {C.~P.}\ \bibnamefont {Koch}}, \ and\ \bibinfo
  {author} {\bibfnamefont {R.}~\bibnamefont {Moszynski}},\ }\href {\doibase
  10.1039/C1CP21196J} {\bibfield  {journal} {\bibinfo  {journal} {Phys. Chem.
  Chem. Phys.}\ }\textbf {\bibinfo {volume} {13}},\ \bibinfo {pages} {18893}
  (\bibinfo {year} {2011})}\BibitemShut {NoStop}%
\bibitem [{\citenamefont {Tomza}\ \emph {et~al.}(2012)\citenamefont {Tomza},
  \citenamefont {Goerz}, \citenamefont {Musia\l{}}, \citenamefont {Moszynski},\
  and\ \citenamefont {Koch}}]{TomzaPRA12}%
  \BibitemOpen
  \bibfield  {author} {\bibinfo {author} {\bibfnamefont {M.}~\bibnamefont
  {Tomza}}, \bibinfo {author} {\bibfnamefont {M.~H.}\ \bibnamefont {Goerz}},
  \bibinfo {author} {\bibfnamefont {M.}~\bibnamefont {Musia\l{}}}, \bibinfo
  {author} {\bibfnamefont {R.}~\bibnamefont {Moszynski}}, \ and\ \bibinfo
  {author} {\bibfnamefont {C.~P.}\ \bibnamefont {Koch}},\ }\href {\doibase
  10.1103/PhysRevA.86.043424} {\bibfield  {journal} {\bibinfo  {journal} {Phys.
  Rev. A}\ }\textbf {\bibinfo {volume} {86}},\ \bibinfo {pages} {043424}
  (\bibinfo {year} {2012})}\BibitemShut {NoStop}%
\bibitem [{\citenamefont {Tomza}\ \emph {et~al.}(2015)\citenamefont {Tomza},
  \citenamefont {Koch},\ and\ \citenamefont {Moszynski}}]{TomzaPRA15a}%
  \BibitemOpen
  \bibfield  {author} {\bibinfo {author} {\bibfnamefont {M.}~\bibnamefont
  {Tomza}}, \bibinfo {author} {\bibfnamefont {C.~P.}\ \bibnamefont {Koch}}, \
  and\ \bibinfo {author} {\bibfnamefont {R.}~\bibnamefont {Moszynski}},\ }\href
  {\doibase 10.1103/PhysRevA.91.042706} {\bibfield  {journal} {\bibinfo
  {journal} {Phys. Rev. A}\ }\textbf {\bibinfo {volume} {91}},\ \bibinfo
  {pages} {042706} (\bibinfo {year} {2015})}\BibitemShut {NoStop}%
\bibitem [{\citenamefont {Werner}\ and\ \citenamefont
  {Knowles}(1988)}]{WernerJCP88}%
  \BibitemOpen
  \bibfield  {author} {\bibinfo {author} {\bibfnamefont {H.}~\bibnamefont
  {Werner}}\ and\ \bibinfo {author} {\bibfnamefont {P.~J.}\ \bibnamefont
  {Knowles}},\ }\href {\doibase http://dx.doi.org/10.1063/1.455556} {\bibfield
  {journal} {\bibinfo  {journal} {J. Chem. Phys.}\ }\textbf {\bibinfo {volume}
  {89}},\ \bibinfo {pages} {5803} (\bibinfo {year} {1988})}\BibitemShut
  {NoStop}%
\bibitem [{\citenamefont {Knowles}\ \emph {et~al.}(1993)\citenamefont
  {Knowles}, \citenamefont {Hampel},\ and\ \citenamefont
  {Werner}}]{KnowlesJCP93}%
  \BibitemOpen
  \bibfield  {author} {\bibinfo {author} {\bibfnamefont {P.~J.}\ \bibnamefont
  {Knowles}}, \bibinfo {author} {\bibfnamefont {C.}~\bibnamefont {Hampel}}, \
  and\ \bibinfo {author} {\bibfnamefont {H.-J.}\ \bibnamefont {Werner}},\
  }\href {\doibase 10.1063/1.465990} {\bibfield  {journal} {\bibinfo  {journal}
  {J. Chem. Phys.}\ }\textbf {\bibinfo {volume} {99}},\ \bibinfo {pages} {5219}
  (\bibinfo {year} {1993})}\BibitemShut {NoStop}%
\bibitem [{\citenamefont {Boys}\ and\ \citenamefont
  {Bernardi}(1970)}]{BoysMP70}%
  \BibitemOpen
  \bibfield  {author} {\bibinfo {author} {\bibfnamefont {S.}~\bibnamefont
  {Boys}}\ and\ \bibinfo {author} {\bibfnamefont {F.}~\bibnamefont
  {Bernardi}},\ }\href {\doibase 10.1080/00268977000101561} {\bibfield
  {journal} {\bibinfo  {journal} {Mol. Phys.}\ }\textbf {\bibinfo {volume}
  {19}},\ \bibinfo {pages} {553} (\bibinfo {year} {1970})}\BibitemShut
  {NoStop}%
\bibitem [{\citenamefont {Woon}\ and\ \citenamefont
  {Dunning~Jr}(1993)}]{WoonJCP93}%
  \BibitemOpen
  \bibfield  {author} {\bibinfo {author} {\bibfnamefont {D.~E.}\ \bibnamefont
  {Woon}}\ and\ \bibinfo {author} {\bibfnamefont {T.~H.}\ \bibnamefont
  {Dunning~Jr}},\ }\href {\doibase 10.1063/1.464303} {\bibfield  {journal}
  {\bibinfo  {journal} {J. Chem. Phys.}\ }\textbf {\bibinfo {volume} {98}},\
  \bibinfo {pages} {1358} (\bibinfo {year} {1993})}\BibitemShut {NoStop}%
\bibitem [{\citenamefont {Dolg}\ and\ \citenamefont {Cao}(2012)}]{DolgCR12}%
  \BibitemOpen
  \bibfield  {author} {\bibinfo {author} {\bibfnamefont {M.}~\bibnamefont
  {Dolg}}\ and\ \bibinfo {author} {\bibfnamefont {X.}~\bibnamefont {Cao}},\
  }\href {\doibase 10.1021/cr2001383} {\bibfield  {journal} {\bibinfo
  {journal} {Chem. Rev.}\ }\textbf {\bibinfo {volume} {112}},\ \bibinfo {pages}
  {403} (\bibinfo {year} {2012})}\BibitemShut {NoStop}%
\bibitem [{\citenamefont {Lim}\ \emph {et~al.}(2005)\citenamefont {Lim},
  \citenamefont {Schwerdtfeger}, \citenamefont {Metz},\ and\ \citenamefont
  {Stoll}}]{LimJCP05}%
  \BibitemOpen
  \bibfield  {author} {\bibinfo {author} {\bibfnamefont {I.~S.}\ \bibnamefont
  {Lim}}, \bibinfo {author} {\bibfnamefont {P.}~\bibnamefont {Schwerdtfeger}},
  \bibinfo {author} {\bibfnamefont {B.}~\bibnamefont {Metz}}, \ and\ \bibinfo
  {author} {\bibfnamefont {H.}~\bibnamefont {Stoll}},\ }\href {\doibase
  10.1063/1.1856451} {\bibfield  {journal} {\bibinfo  {journal} {J. Chem.
  Phys.}\ }\textbf {\bibinfo {volume} {122}},\ \bibinfo {pages} {104103}
  (\bibinfo {year} {2005})}\BibitemShut {NoStop}%
\bibitem [{\citenamefont {Lim}\ \emph {et~al.}(2006)\citenamefont {Lim},
  \citenamefont {Stoll},\ and\ \citenamefont {Schwerdtfeger}}]{LimJCP06}%
  \BibitemOpen
  \bibfield  {author} {\bibinfo {author} {\bibfnamefont {I.~S.}\ \bibnamefont
  {Lim}}, \bibinfo {author} {\bibfnamefont {H.}~\bibnamefont {Stoll}}, \ and\
  \bibinfo {author} {\bibfnamefont {P.}~\bibnamefont {Schwerdtfeger}},\ }\href
  {\doibase 10.1063/1.2148945} {\bibfield  {journal} {\bibinfo  {journal} {J.
  Chem. Phys.}\ }\textbf {\bibinfo {volume} {124}},\ \bibinfo {pages} {034107}
  (\bibinfo {year} {2006})}\BibitemShut {NoStop}%
\bibitem [{\citenamefont {Tomza}\ \emph {et~al.}(2013)\citenamefont {Tomza},
  \citenamefont {Skomorowski}, \citenamefont {Musial}, \citenamefont
  {Gonzalez~Ferez}, \citenamefont {Koch},\ and\ \citenamefont
  {Moszynski}}]{TomzaMP13}%
  \BibitemOpen
  \bibfield  {author} {\bibinfo {author} {\bibfnamefont {M.}~\bibnamefont
  {Tomza}}, \bibinfo {author} {\bibfnamefont {W.}~\bibnamefont {Skomorowski}},
  \bibinfo {author} {\bibfnamefont {M.}~\bibnamefont {Musial}}, \bibinfo
  {author} {\bibfnamefont {R.}~\bibnamefont {Gonzalez~Ferez}}, \bibinfo
  {author} {\bibfnamefont {C.~P.}\ \bibnamefont {Koch}}, \ and\ \bibinfo
  {author} {\bibfnamefont {R.}~\bibnamefont {Moszynski}},\ }\href {\doibase
  10.1080/00268976.2013.793835} {\bibfield  {journal} {\bibinfo  {journal}
  {Mol. Phys.}\ }\textbf {\bibinfo {volume} {111}},\ \bibinfo {pages} {1781}
  (\bibinfo {year} {2013})}\BibitemShut {NoStop}%
\bibitem [{\citenamefont {Tao}\ and\ \citenamefont {Pan}(1992)}]{midbond}%
  \BibitemOpen
  \bibfield  {author} {\bibinfo {author} {\bibfnamefont {F.-M.}\ \bibnamefont
  {Tao}}\ and\ \bibinfo {author} {\bibfnamefont {Y.-K.}\ \bibnamefont {Pan}},\
  }\href {\doibase 10.1063/1.463852} {\bibfield  {journal} {\bibinfo  {journal}
  {J. Chem. Phys.}\ }\textbf {\bibinfo {volume} {97}},\ \bibinfo {pages} {4989}
  (\bibinfo {year} {1992})}\BibitemShut {NoStop}%
\bibitem [{\citenamefont {Werner}\ \emph {et~al.}(2012)\citenamefont {Werner},
  \citenamefont {Knowles}, \citenamefont {Knizia}, \citenamefont {Manby},
  \citenamefont {{Sch\"{u}tz}}, \citenamefont {Celani}, \citenamefont {Korona},
  \citenamefont {Lindh}, \citenamefont {Mitrushenkov}, \citenamefont {Rauhut},
  \citenamefont {Shamasundar}, \citenamefont {Adler}, \citenamefont {Amos},
  \citenamefont {Bernhardsson}, \citenamefont {Berning}, \citenamefont
  {Cooper}, \citenamefont {Deegan}, \citenamefont {Dobbyn}, \citenamefont
  {Eckert}, \citenamefont {Goll}, \citenamefont {Hampel}, \citenamefont
  {Hesselmann}, \citenamefont {Hetzer}, \citenamefont {Hrenar}, \citenamefont
  {Jansen}, \citenamefont {K\"oppl}, \citenamefont {Liu}, \citenamefont
  {Lloyd}, \citenamefont {Mata}, \citenamefont {May}, \citenamefont
  {McNicholas}, \citenamefont {Meyer}, \citenamefont {Mura}, \citenamefont
  {Nicklass}, \citenamefont {O'Neill}, \citenamefont {Palmieri}, \citenamefont
  {Peng}, \citenamefont {Pfl\"uger}, \citenamefont {Pitzer}, \citenamefont
  {Reiher}, \citenamefont {Shiozaki}, \citenamefont {Stoll}, \citenamefont
  {Stone}, \citenamefont {Tarroni}, \citenamefont {Thorsteinsson},\ and\
  \citenamefont {Wang}}]{Molpro}%
  \BibitemOpen
  \bibfield  {author} {\bibinfo {author} {\bibfnamefont {H.-J.}\ \bibnamefont
  {Werner}}, \bibinfo {author} {\bibfnamefont {P.~J.}\ \bibnamefont {Knowles}},
  \bibinfo {author} {\bibfnamefont {G.}~\bibnamefont {Knizia}}, \bibinfo
  {author} {\bibfnamefont {F.~R.}\ \bibnamefont {Manby}}, \bibinfo {author}
  {\bibfnamefont {M.}~\bibnamefont {{Sch\"{u}tz}}}, \bibinfo {author}
  {\bibfnamefont {P.}~\bibnamefont {Celani}}, \bibinfo {author} {\bibfnamefont
  {T.}~\bibnamefont {Korona}}, \bibinfo {author} {\bibfnamefont
  {R.}~\bibnamefont {Lindh}}, \bibinfo {author} {\bibfnamefont
  {A.}~\bibnamefont {Mitrushenkov}}, \bibinfo {author} {\bibfnamefont
  {G.}~\bibnamefont {Rauhut}}, \bibinfo {author} {\bibfnamefont {K.~R.}\
  \bibnamefont {Shamasundar}}, \bibinfo {author} {\bibfnamefont {T.~B.}\
  \bibnamefont {Adler}}, \bibinfo {author} {\bibfnamefont {R.~D.}\ \bibnamefont
  {Amos}}, \bibinfo {author} {\bibfnamefont {A.}~\bibnamefont {Bernhardsson}},
  \bibinfo {author} {\bibfnamefont {A.}~\bibnamefont {Berning}}, \bibinfo
  {author} {\bibfnamefont {D.~L.}\ \bibnamefont {Cooper}}, \bibinfo {author}
  {\bibfnamefont {M.~J.~O.}\ \bibnamefont {Deegan}}, \bibinfo {author}
  {\bibfnamefont {A.~J.}\ \bibnamefont {Dobbyn}}, \bibinfo {author}
  {\bibfnamefont {F.}~\bibnamefont {Eckert}}, \bibinfo {author} {\bibfnamefont
  {E.}~\bibnamefont {Goll}}, \bibinfo {author} {\bibfnamefont {C.}~\bibnamefont
  {Hampel}}, \bibinfo {author} {\bibfnamefont {A.}~\bibnamefont {Hesselmann}},
  \bibinfo {author} {\bibfnamefont {G.}~\bibnamefont {Hetzer}}, \bibinfo
  {author} {\bibfnamefont {T.}~\bibnamefont {Hrenar}}, \bibinfo {author}
  {\bibfnamefont {G.}~\bibnamefont {Jansen}}, \bibinfo {author} {\bibfnamefont
  {C.}~\bibnamefont {K\"oppl}}, \bibinfo {author} {\bibfnamefont
  {Y.}~\bibnamefont {Liu}}, \bibinfo {author} {\bibfnamefont {A.~W.}\
  \bibnamefont {Lloyd}}, \bibinfo {author} {\bibfnamefont {R.~A.}\ \bibnamefont
  {Mata}}, \bibinfo {author} {\bibfnamefont {A.~J.}\ \bibnamefont {May}},
  \bibinfo {author} {\bibfnamefont {S.~J.}\ \bibnamefont {McNicholas}},
  \bibinfo {author} {\bibfnamefont {W.}~\bibnamefont {Meyer}}, \bibinfo
  {author} {\bibfnamefont {M.~E.}\ \bibnamefont {Mura}}, \bibinfo {author}
  {\bibfnamefont {A.}~\bibnamefont {Nicklass}}, \bibinfo {author}
  {\bibfnamefont {D.~P.}\ \bibnamefont {O'Neill}}, \bibinfo {author}
  {\bibfnamefont {P.}~\bibnamefont {Palmieri}}, \bibinfo {author}
  {\bibfnamefont {D.}~\bibnamefont {Peng}}, \bibinfo {author} {\bibfnamefont
  {K.}~\bibnamefont {Pfl\"uger}}, \bibinfo {author} {\bibfnamefont
  {R.}~\bibnamefont {Pitzer}}, \bibinfo {author} {\bibfnamefont
  {M.}~\bibnamefont {Reiher}}, \bibinfo {author} {\bibfnamefont
  {T.}~\bibnamefont {Shiozaki}}, \bibinfo {author} {\bibfnamefont
  {H.}~\bibnamefont {Stoll}}, \bibinfo {author} {\bibfnamefont {A.~J.}\
  \bibnamefont {Stone}}, \bibinfo {author} {\bibfnamefont {R.}~\bibnamefont
  {Tarroni}}, \bibinfo {author} {\bibfnamefont {T.}~\bibnamefont
  {Thorsteinsson}}, \ and\ \bibinfo {author} {\bibfnamefont {M.}~\bibnamefont
  {Wang}},\ }\href@noop {} {\enquote {\bibinfo {title} {Molpro, version 2012.1,
  a package of ab initio programs},}\ } (\bibinfo {year} {2012}),\ \bibinfo
  {note} {see http://www.molpro.net}\BibitemShut {NoStop}%
\bibitem [{\citenamefont {Dyall}\ and\ \citenamefont
  {F{\ae}gri~Jr}(2007)}]{Dyall2007}%
  \BibitemOpen
  \bibfield  {author} {\bibinfo {author} {\bibfnamefont {K.~G.}\ \bibnamefont
  {Dyall}}\ and\ \bibinfo {author} {\bibfnamefont {K.}~\bibnamefont
  {F{\ae}gri~Jr}},\ }\href@noop {} {\emph {\bibinfo {title} {Introduction to
  relativistic quantum chemistry}}}\ (\bibinfo  {publisher} {Oxford University
  Press},\ \bibinfo {year} {2007})\BibitemShut {NoStop}%
\bibitem [{nis()}]{nist}%
  \BibitemOpen
  \href@noop {} {}\bibinfo {note} {{NIST Atomic Spectra Database
  http://physics.nist.gov/PhysRefData/ASD}}\BibitemShut {NoStop}%
\bibitem [{\citenamefont {Jeziorski}\ \emph {et~al.}(1994)\citenamefont
  {Jeziorski}, \citenamefont {Moszynski},\ and\ \citenamefont
  {Szalewicz}}]{JeziorskiCR94}%
  \BibitemOpen
  \bibfield  {author} {\bibinfo {author} {\bibfnamefont {B.}~\bibnamefont
  {Jeziorski}}, \bibinfo {author} {\bibfnamefont {R.}~\bibnamefont
  {Moszynski}}, \ and\ \bibinfo {author} {\bibfnamefont {K.}~\bibnamefont
  {Szalewicz}},\ }\href {\doibase 10.1021/cr00031a008} {\bibfield  {journal}
  {\bibinfo  {journal} {Chem. Rev.}\ }\textbf {\bibinfo {volume} {94}},\
  \bibinfo {pages} {1887} (\bibinfo {year} {1994})}\BibitemShut {NoStop}%
\bibitem [{\citenamefont {Derevianko}\ \emph {et~al.}(2010)\citenamefont
  {Derevianko}, \citenamefont {Porsev},\ and\ \citenamefont
  {Babb}}]{DerevienkoADNDT10}%
  \BibitemOpen
  \bibfield  {author} {\bibinfo {author} {\bibfnamefont {A.}~\bibnamefont
  {Derevianko}}, \bibinfo {author} {\bibfnamefont {S.~G.}\ \bibnamefont
  {Porsev}}, \ and\ \bibinfo {author} {\bibfnamefont {J.~F.}\ \bibnamefont
  {Babb}},\ }\href {\doibase 10.1016/j.adt.2009.12.002} {\bibfield  {journal}
  {\bibinfo  {journal} {{At. Data Nucl. Data Tables}}\ }\textbf {\bibinfo
  {volume} {{96}}},\ \bibinfo {pages} {{323}} (\bibinfo {year}
  {{2010}})}\BibitemShut {NoStop}%
\bibitem [{\citenamefont {Korona}\ \emph {et~al.}(2006)\citenamefont {Korona},
  \citenamefont {Przybytek},\ and\ \citenamefont {Jeziorski}}]{KoronaMP06}%
  \BibitemOpen
  \bibfield  {author} {\bibinfo {author} {\bibfnamefont {T.}~\bibnamefont
  {Korona}}, \bibinfo {author} {\bibfnamefont {M.}~\bibnamefont {Przybytek}}, \
  and\ \bibinfo {author} {\bibfnamefont {B.}~\bibnamefont {Jeziorski}},\ }\href
  {\doibase 10.1080/00268970600673975} {\bibfield  {journal} {\bibinfo
  {journal} {{Mol. Phys.}}\ }\textbf {\bibinfo {volume} {{104}}},\ \bibinfo
  {pages} {{2303}} (\bibinfo {year} {{2006}})}\BibitemShut {NoStop}%
\bibitem [{\citenamefont {Tomza}\ \emph {et~al.}(2014)\citenamefont {Tomza},
  \citenamefont {Gonz\'alez-F\'erez}, \citenamefont {Koch},\ and\ \citenamefont
  {Moszynski}}]{TomzaPRL14}%
  \BibitemOpen
  \bibfield  {author} {\bibinfo {author} {\bibfnamefont {M.}~\bibnamefont
  {Tomza}}, \bibinfo {author} {\bibfnamefont {R.}~\bibnamefont
  {Gonz\'alez-F\'erez}}, \bibinfo {author} {\bibfnamefont {C.~P.}\ \bibnamefont
  {Koch}}, \ and\ \bibinfo {author} {\bibfnamefont {R.}~\bibnamefont
  {Moszynski}},\ }\href {\doibase 10.1103/PhysRevLett.112.113201} {\bibfield
  {journal} {\bibinfo  {journal} {Phys. Rev. Lett.}\ }\textbf {\bibinfo
  {volume} {112}},\ \bibinfo {pages} {113201} (\bibinfo {year}
  {2014})}\BibitemShut {NoStop}%
\bibitem [{\citenamefont {Johnson}(1978)}]{JohnsonJCP78}%
  \BibitemOpen
  \bibfield  {author} {\bibinfo {author} {\bibfnamefont {B.~R.}\ \bibnamefont
  {Johnson}},\ }\href {\doibase 10.1063/1.436421} {\bibfield  {journal}
  {\bibinfo  {journal} {J. Chem. Phys.}\ }\textbf {\bibinfo {volume} {69}},\
  \bibinfo {pages} {4678} (\bibinfo {year} {1978})}\BibitemShut {NoStop}%
\bibitem [{\citenamefont {Sayfutyarova}\ \emph {et~al.}(2013)\citenamefont
  {Sayfutyarova}, \citenamefont {Buchachenko}, \citenamefont {Yakovleva},\ and\
  \citenamefont {Belyaev}}]{SayfutyarovaPRA13}%
  \BibitemOpen
  \bibfield  {author} {\bibinfo {author} {\bibfnamefont {E.~R.}\ \bibnamefont
  {Sayfutyarova}}, \bibinfo {author} {\bibfnamefont {A.~A.}\ \bibnamefont
  {Buchachenko}}, \bibinfo {author} {\bibfnamefont {S.~A.}\ \bibnamefont
  {Yakovleva}}, \ and\ \bibinfo {author} {\bibfnamefont {A.~K.}\ \bibnamefont
  {Belyaev}},\ }\href {\doibase 10.1103/PhysRevA.87.052717} {\bibfield
  {journal} {\bibinfo  {journal} {Phys. Rev. A}\ }\textbf {\bibinfo {volume}
  {87}},\ \bibinfo {pages} {052717} (\bibinfo {year} {2013})}\BibitemShut
  {NoStop}%
\bibitem [{\citenamefont {Li}\ \emph {et~al.}(2019)\citenamefont {Li},
  \citenamefont {Mills}, \citenamefont {Puri}, \citenamefont {Petrov},
  \citenamefont {Hudson},\ and\ \citenamefont {Kotochigova}}]{LiPRA19}%
  \BibitemOpen
  \bibfield  {author} {\bibinfo {author} {\bibfnamefont {M.}~\bibnamefont
  {Li}}, \bibinfo {author} {\bibfnamefont {M.}~\bibnamefont {Mills}}, \bibinfo
  {author} {\bibfnamefont {P.}~\bibnamefont {Puri}}, \bibinfo {author}
  {\bibfnamefont {A.}~\bibnamefont {Petrov}}, \bibinfo {author} {\bibfnamefont
  {E.~R.}\ \bibnamefont {Hudson}}, \ and\ \bibinfo {author} {\bibfnamefont
  {S.}~\bibnamefont {Kotochigova}},\ }\href {\doibase
  10.1103/PhysRevA.99.062706} {\bibfield  {journal} {\bibinfo  {journal} {Phys.
  Rev. A}\ }\textbf {\bibinfo {volume} {99}},\ \bibinfo {pages} {062706}
  (\bibinfo {year} {2019})}\BibitemShut {NoStop}%
\bibitem [{\citenamefont {Mitroy}\ \emph {et~al.}(2010)\citenamefont {Mitroy},
  \citenamefont {Safronova},\ and\ \citenamefont {Clark}}]{MitroyJPB10}%
  \BibitemOpen
  \bibfield  {author} {\bibinfo {author} {\bibfnamefont {J.}~\bibnamefont
  {Mitroy}}, \bibinfo {author} {\bibfnamefont {M.~S.}\ \bibnamefont
  {Safronova}}, \ and\ \bibinfo {author} {\bibfnamefont {C.~W.}\ \bibnamefont
  {Clark}},\ }\href {\doibase 10.1088/0953-4075/43/20/202001} {\bibfield
  {journal} {\bibinfo  {journal} {J. Phys. B: At. Mol. Opt. Phys.}\ }\textbf
  {\bibinfo {volume} {43}},\ \bibinfo {pages} {202001} (\bibinfo {year}
  {2010})}\BibitemShut {NoStop}%
\bibitem [{\citenamefont {Krych}\ \emph {et~al.}(2011)\citenamefont {Krych},
  \citenamefont {Skomorowski}, \citenamefont {Paw\l{}owski}, \citenamefont
  {Moszynski},\ and\ \citenamefont {Idziaszek}}]{KrychPRA11}%
  \BibitemOpen
  \bibfield  {author} {\bibinfo {author} {\bibfnamefont {M.}~\bibnamefont
  {Krych}}, \bibinfo {author} {\bibfnamefont {W.}~\bibnamefont {Skomorowski}},
  \bibinfo {author} {\bibfnamefont {F.}~\bibnamefont {Paw\l{}owski}}, \bibinfo
  {author} {\bibfnamefont {R.}~\bibnamefont {Moszynski}}, \ and\ \bibinfo
  {author} {\bibfnamefont {Z.}~\bibnamefont {Idziaszek}},\ }\href {\doibase
  10.1103/PhysRevA.83.032723} {\bibfield  {journal} {\bibinfo  {journal} {Phys.
  Rev. A}\ }\textbf {\bibinfo {volume} {83}},\ \bibinfo {pages} {032723}
  (\bibinfo {year} {2011})}\BibitemShut {NoStop}%
\bibitem [{\citenamefont {Idziaszek}\ \emph {et~al.}(2011)\citenamefont
  {Idziaszek}, \citenamefont {Simoni}, \citenamefont {Calarco},\ and\
  \citenamefont {Julienne}}]{IdziaszekNJP11}%
  \BibitemOpen
  \bibfield  {author} {\bibinfo {author} {\bibfnamefont {Z.}~\bibnamefont
  {Idziaszek}}, \bibinfo {author} {\bibfnamefont {A.}~\bibnamefont {Simoni}},
  \bibinfo {author} {\bibfnamefont {T.}~\bibnamefont {Calarco}}, \ and\
  \bibinfo {author} {\bibfnamefont {P.~S.}\ \bibnamefont {Julienne}},\ }\href
  {\doibase 10.1088/1367-2630/13/8/083005} {\bibfield  {journal} {\bibinfo
  {journal} {New J. Phys.}\ }\textbf {\bibinfo {volume} {13}},\ \bibinfo
  {pages} {083005} (\bibinfo {year} {2011})}\BibitemShut {NoStop}%
\bibitem [{\citenamefont {da~Silva~Jr}\ \emph {et~al.}(2015)\citenamefont
  {da~Silva~Jr}, \citenamefont {Raoult}, \citenamefont {Aymar},\ and\
  \citenamefont {Dulieu}}]{daSilvaNJP2015}%
  \BibitemOpen
  \bibfield  {author} {\bibinfo {author} {\bibfnamefont {H.}~\bibnamefont
  {da~Silva~Jr}}, \bibinfo {author} {\bibfnamefont {M.}~\bibnamefont {Raoult}},
  \bibinfo {author} {\bibfnamefont {M.}~\bibnamefont {Aymar}}, \ and\ \bibinfo
  {author} {\bibfnamefont {O.}~\bibnamefont {Dulieu}},\ }\href {\doibase
  10.1088/1367-2630/17/4/045015} {\bibfield  {journal} {\bibinfo  {journal}
  {New J. Phys.}\ }\textbf {\bibinfo {volume} {17}},\ \bibinfo {pages} {045015}
  (\bibinfo {year} {2015})}\BibitemShut {NoStop}%
\bibitem [{\citenamefont {Rouse}\ and\ \citenamefont
  {Willitsch}(2017)}]{RousePRL17}%
  \BibitemOpen
  \bibfield  {author} {\bibinfo {author} {\bibfnamefont {I.}~\bibnamefont
  {Rouse}}\ and\ \bibinfo {author} {\bibfnamefont {S.}~\bibnamefont
  {Willitsch}},\ }\href {\doibase 10.1103/PhysRevLett.118.143401} {\bibfield
  {journal} {\bibinfo  {journal} {Phys. Rev. Lett.}\ }\textbf {\bibinfo
  {volume} {118}},\ \bibinfo {pages} {143401} (\bibinfo {year}
  {2017})}\BibitemShut {NoStop}%
\bibitem [{\citenamefont {Levine}(2009)}]{Levine09}%
  \BibitemOpen
  \bibfield  {author} {\bibinfo {author} {\bibfnamefont {R.~D.}\ \bibnamefont
  {Levine}},\ }\href@noop {} {\emph {\bibinfo {title} {Molecular reaction
  dynamics}}}\ (\bibinfo  {publisher} {Cambridge University Press},\ \bibinfo
  {year} {2009})\BibitemShut {NoStop}%
\bibitem [{\citenamefont {Belyaev}\ and\ \citenamefont
  {Lebedev}(2011)}]{BelyaevPRA11}%
  \BibitemOpen
  \bibfield  {author} {\bibinfo {author} {\bibfnamefont {A.~K.}\ \bibnamefont
  {Belyaev}}\ and\ \bibinfo {author} {\bibfnamefont {O.~V.}\ \bibnamefont
  {Lebedev}},\ }\href {\doibase 10.1103/PhysRevA.84.014701} {\bibfield
  {journal} {\bibinfo  {journal} {Phys. Rev. A}\ }\textbf {\bibinfo {volume}
  {84}},\ \bibinfo {pages} {014701} (\bibinfo {year} {2011})}\BibitemShut
  {NoStop}%
\bibitem [{\citenamefont {Brewer}\ \emph
  {et~al.}(2019{\natexlab{b}})\citenamefont {Brewer}, \citenamefont {Chen},
  \citenamefont {Beloy}, \citenamefont {Hankin}, \citenamefont {Clements},
  \citenamefont {Chou}, \citenamefont {McGrew}, \citenamefont {Zhang},
  \citenamefont {Fasano}, \citenamefont {Nicolodi}, \citenamefont {Leopardi},
  \citenamefont {Fortier}, \citenamefont {Diddams}, \citenamefont {Ludlow},
  \citenamefont {Wineland}, \citenamefont {Leibrandt},\ and\ \citenamefont
  {Hume}}]{Brewer2019b}%
  \BibitemOpen
  \bibfield  {author} {\bibinfo {author} {\bibfnamefont {S.~M.}\ \bibnamefont
  {Brewer}}, \bibinfo {author} {\bibfnamefont {J.-S.}\ \bibnamefont {Chen}},
  \bibinfo {author} {\bibfnamefont {K.}~\bibnamefont {Beloy}}, \bibinfo
  {author} {\bibfnamefont {A.~M.}\ \bibnamefont {Hankin}}, \bibinfo {author}
  {\bibfnamefont {E.~R.}\ \bibnamefont {Clements}}, \bibinfo {author}
  {\bibfnamefont {C.~W.}\ \bibnamefont {Chou}}, \bibinfo {author}
  {\bibfnamefont {W.~F.}\ \bibnamefont {McGrew}}, \bibinfo {author}
  {\bibfnamefont {X.}~\bibnamefont {Zhang}}, \bibinfo {author} {\bibfnamefont
  {R.~J.}\ \bibnamefont {Fasano}}, \bibinfo {author} {\bibfnamefont
  {D.}~\bibnamefont {Nicolodi}}, \bibinfo {author} {\bibfnamefont
  {H.}~\bibnamefont {Leopardi}}, \bibinfo {author} {\bibfnamefont {T.~M.}\
  \bibnamefont {Fortier}}, \bibinfo {author} {\bibfnamefont {S.~A.}\
  \bibnamefont {Diddams}}, \bibinfo {author} {\bibfnamefont {A.~D.}\
  \bibnamefont {Ludlow}}, \bibinfo {author} {\bibfnamefont {D.~J.}\
  \bibnamefont {Wineland}}, \bibinfo {author} {\bibfnamefont {D.~R.}\
  \bibnamefont {Leibrandt}}, \ and\ \bibinfo {author} {\bibfnamefont {D.~B.}\
  \bibnamefont {Hume}},\ }\href {\doibase 10.1103/PhysRevA.100.013409}
  {\bibfield  {journal} {\bibinfo  {journal} {Phys. Rev. A}\ }\textbf {\bibinfo
  {volume} {100}},\ \bibinfo {pages} {013409} (\bibinfo {year}
  {2019}{\natexlab{b}})}\BibitemShut {NoStop}%
\bibitem [{\citenamefont {Guggemos}\ \emph {et~al.}(2019)\citenamefont
  {Guggemos}, \citenamefont {Guevara-Bertsch}, \citenamefont {Heinrich},
  \citenamefont {Herrera-Sancho}, \citenamefont {Colombe}, \citenamefont
  {Blatt},\ and\ \citenamefont {Roos}}]{Guggemos2019}%
  \BibitemOpen
  \bibfield  {author} {\bibinfo {author} {\bibfnamefont {M.}~\bibnamefont
  {Guggemos}}, \bibinfo {author} {\bibfnamefont {M.}~\bibnamefont
  {Guevara-Bertsch}}, \bibinfo {author} {\bibfnamefont {D.}~\bibnamefont
  {Heinrich}}, \bibinfo {author} {\bibfnamefont {O.~A.}\ \bibnamefont
  {Herrera-Sancho}}, \bibinfo {author} {\bibfnamefont {Y.}~\bibnamefont
  {Colombe}}, \bibinfo {author} {\bibfnamefont {R.}~\bibnamefont {Blatt}}, \
  and\ \bibinfo {author} {\bibfnamefont {C.~F.}\ \bibnamefont {Roos}},\ }\href
  {\doibase 10.1088/1367-2630/ab447a} {\bibfield  {journal} {\bibinfo
  {journal} {New J. Phys.}\ }\textbf {\bibinfo {volume} {21}},\ \bibinfo
  {pages} {103003} (\bibinfo {year} {2019})}\BibitemShut {NoStop}%
\bibitem [{\citenamefont {Meir}\ \emph {et~al.}(2018)\citenamefont {Meir},
  \citenamefont {Pinkas}, \citenamefont {Sikorsky}, \citenamefont {Ben-shlomi},
  \citenamefont {Akerman},\ and\ \citenamefont {Ozeri}}]{MeirPRL18}%
  \BibitemOpen
  \bibfield  {author} {\bibinfo {author} {\bibfnamefont {Z.}~\bibnamefont
  {Meir}}, \bibinfo {author} {\bibfnamefont {M.}~\bibnamefont {Pinkas}},
  \bibinfo {author} {\bibfnamefont {T.}~\bibnamefont {Sikorsky}}, \bibinfo
  {author} {\bibfnamefont {R.}~\bibnamefont {Ben-shlomi}}, \bibinfo {author}
  {\bibfnamefont {N.}~\bibnamefont {Akerman}}, \ and\ \bibinfo {author}
  {\bibfnamefont {R.}~\bibnamefont {Ozeri}},\ }\href {\doibase
  10.1103/PhysRevLett.121.053402} {\bibfield  {journal} {\bibinfo  {journal}
  {Phys. Rev. Lett.}\ }\textbf {\bibinfo {volume} {121}},\ \bibinfo {pages}
  {053402} (\bibinfo {year} {2018})}\BibitemShut {NoStop}%
\bibitem [{\citenamefont {Cetina}\ \emph {et~al.}(2012)\citenamefont {Cetina},
  \citenamefont {Grier},\ and\ \citenamefont {Vuleti\ifmmode~\acute{c}\else
  \'{c}\fi{}}}]{CetinaPRL12}%
  \BibitemOpen
  \bibfield  {author} {\bibinfo {author} {\bibfnamefont {M.}~\bibnamefont
  {Cetina}}, \bibinfo {author} {\bibfnamefont {A.~T.}\ \bibnamefont {Grier}}, \
  and\ \bibinfo {author} {\bibfnamefont {V.}~\bibnamefont
  {Vuleti\ifmmode~\acute{c}\else \'{c}\fi{}}},\ }\href {\doibase
  10.1103/PhysRevLett.109.253201} {\bibfield  {journal} {\bibinfo  {journal}
  {Phys. Rev. Lett.}\ }\textbf {\bibinfo {volume} {109}},\ \bibinfo {pages}
  {253201} (\bibinfo {year} {2012})}\BibitemShut {NoStop}%
\bibitem [{\citenamefont {Schmidt}\ \emph {et~al.}(2019)\citenamefont
  {Schmidt}, \citenamefont {Weckesser}, \citenamefont {Thielemann},
  \citenamefont {Schaetz},\ and\ \citenamefont {Karpa}}]{Schmidt2019}%
  \BibitemOpen
  \bibfield  {author} {\bibinfo {author} {\bibfnamefont {J.}~\bibnamefont
  {Schmidt}}, \bibinfo {author} {\bibfnamefont {P.}~\bibnamefont {Weckesser}},
  \bibinfo {author} {\bibfnamefont {F.}~\bibnamefont {Thielemann}}, \bibinfo
  {author} {\bibfnamefont {T.}~\bibnamefont {Schaetz}}, \ and\ \bibinfo
  {author} {\bibfnamefont {L.}~\bibnamefont {Karpa}},\ }\href@noop {}
  {\bibfield  {journal} {\bibinfo  {journal} {arXiv preprint arXiv:1909.08352}\
  } (\bibinfo {year} {2019})}\BibitemShut {NoStop}%
\bibitem [{\citenamefont {Schneider}\ \emph {et~al.}(2010)\citenamefont
  {Schneider}, \citenamefont {Enderlein}, \citenamefont {Huber},\ and\
  \citenamefont {Schaetz}}]{SchneiderNatPhot10}%
  \BibitemOpen
  \bibfield  {author} {\bibinfo {author} {\bibfnamefont {C.}~\bibnamefont
  {Schneider}}, \bibinfo {author} {\bibfnamefont {M.}~\bibnamefont
  {Enderlein}}, \bibinfo {author} {\bibfnamefont {T.}~\bibnamefont {Huber}}, \
  and\ \bibinfo {author} {\bibfnamefont {T.}~\bibnamefont {Schaetz}},\ }\href
  {\doibase 10.1038/nphoton.2010.236} {\bibfield  {journal} {\bibinfo
  {journal} {Nat. Photon.}\ }\textbf {\bibinfo {volume} {4}},\ \bibinfo {pages}
  {772} (\bibinfo {year} {2010})}\BibitemShut {NoStop}%
\bibitem [{\citenamefont {Gacesa}\ \emph {et~al.}(2016)\citenamefont {Gacesa},
  \citenamefont {Montgomery}, \citenamefont {Michels},\ and\ \citenamefont
  {C\^ot\'e}}]{GacesaPRA16}%
  \BibitemOpen
  \bibfield  {author} {\bibinfo {author} {\bibfnamefont {M.}~\bibnamefont
  {Gacesa}}, \bibinfo {author} {\bibfnamefont {J.~A.}\ \bibnamefont
  {Montgomery}}, \bibinfo {author} {\bibfnamefont {H.~H.}\ \bibnamefont
  {Michels}}, \ and\ \bibinfo {author} {\bibfnamefont {R.}~\bibnamefont
  {C\^ot\'e}},\ }\href {\doibase 10.1103/PhysRevA.94.013407} {\bibfield
  {journal} {\bibinfo  {journal} {Phys. Rev. A}\ }\textbf {\bibinfo {volume}
  {94}},\ \bibinfo {pages} {013407} (\bibinfo {year} {2016})}\BibitemShut
  {NoStop}%
\bibitem [{\citenamefont {Vitanov}\ \emph {et~al.}(2017)\citenamefont
  {Vitanov}, \citenamefont {Rangelov}, \citenamefont {Shore},\ and\
  \citenamefont {Bergmann}}]{VitanovRMP17}%
  \BibitemOpen
  \bibfield  {author} {\bibinfo {author} {\bibfnamefont {N.~V.}\ \bibnamefont
  {Vitanov}}, \bibinfo {author} {\bibfnamefont {A.~A.}\ \bibnamefont
  {Rangelov}}, \bibinfo {author} {\bibfnamefont {B.~W.}\ \bibnamefont {Shore}},
  \ and\ \bibinfo {author} {\bibfnamefont {K.}~\bibnamefont {Bergmann}},\
  }\href {\doibase 10.1103/RevModPhys.89.015006} {\bibfield  {journal}
  {\bibinfo  {journal} {Rev. Mod. Phys.}\ }\textbf {\bibinfo {volume} {89}},\
  \bibinfo {pages} {015006} (\bibinfo {year} {2017})}\BibitemShut {NoStop}%
\bibitem [{\citenamefont {Chen}\ \emph {et~al.}(2011)\citenamefont {Chen},
  \citenamefont {Schowalter}, \citenamefont {Kotochigova}, \citenamefont
  {Petrov}, \citenamefont {Rellergert}, \citenamefont {Sullivan},\ and\
  \citenamefont {Hudson}}]{KuangPRA11}%
  \BibitemOpen
  \bibfield  {author} {\bibinfo {author} {\bibfnamefont {K.}~\bibnamefont
  {Chen}}, \bibinfo {author} {\bibfnamefont {S.~J.}\ \bibnamefont
  {Schowalter}}, \bibinfo {author} {\bibfnamefont {S.}~\bibnamefont
  {Kotochigova}}, \bibinfo {author} {\bibfnamefont {A.}~\bibnamefont {Petrov}},
  \bibinfo {author} {\bibfnamefont {W.~G.}\ \bibnamefont {Rellergert}},
  \bibinfo {author} {\bibfnamefont {S.~T.}\ \bibnamefont {Sullivan}}, \ and\
  \bibinfo {author} {\bibfnamefont {E.~R.}\ \bibnamefont {Hudson}},\ }\href
  {\doibase 10.1103/PhysRevA.83.030501} {\bibfield  {journal} {\bibinfo
  {journal} {Phys. Rev. A}\ }\textbf {\bibinfo {volume} {83}},\ \bibinfo
  {pages} {030501} (\bibinfo {year} {2011})}\BibitemShut {NoStop}%
\bibitem [{\citenamefont {Wolf}\ \emph {et~al.}(2016)\citenamefont {Wolf},
  \citenamefont {Wan}, \citenamefont {Heip}, \citenamefont {Gebert},
  \citenamefont {Shi},\ and\ \citenamefont {Schmidt}}]{WolfNature16}%
  \BibitemOpen
  \bibfield  {author} {\bibinfo {author} {\bibfnamefont {F.}~\bibnamefont
  {Wolf}}, \bibinfo {author} {\bibfnamefont {Y.}~\bibnamefont {Wan}}, \bibinfo
  {author} {\bibfnamefont {J.~C.}\ \bibnamefont {Heip}}, \bibinfo {author}
  {\bibfnamefont {F.}~\bibnamefont {Gebert}}, \bibinfo {author} {\bibfnamefont
  {C.}~\bibnamefont {Shi}}, \ and\ \bibinfo {author} {\bibfnamefont {P.~O.}\
  \bibnamefont {Schmidt}},\ }\href {\doibase 10.1038/nature16513} {\bibfield
  {journal} {\bibinfo  {journal} {Nature}\ }\textbf {\bibinfo {volume} {530}},\
  \bibinfo {pages} {457} (\bibinfo {year} {2016})}\BibitemShut {NoStop}%
\bibitem [{\citenamefont {Gibble}(2013)}]{GibblePRL13}%
  \BibitemOpen
  \bibfield  {author} {\bibinfo {author} {\bibfnamefont {K.}~\bibnamefont
  {Gibble}},\ }\href {\doibase 10.1103/PhysRevLett.110.180802} {\bibfield
  {journal} {\bibinfo  {journal} {Phys. Rev. Lett.}\ }\textbf {\bibinfo
  {volume} {110}},\ \bibinfo {pages} {180802} (\bibinfo {year}
  {2013})}\BibitemShut {NoStop}%
\bibitem [{\citenamefont {Vutha}\ \emph {et~al.}(2017)\citenamefont {Vutha},
  \citenamefont {Kirchner},\ and\ \citenamefont {Dub\'e}}]{VuthaPRA17}%
  \BibitemOpen
  \bibfield  {author} {\bibinfo {author} {\bibfnamefont {A.~C.}\ \bibnamefont
  {Vutha}}, \bibinfo {author} {\bibfnamefont {T.}~\bibnamefont {Kirchner}}, \
  and\ \bibinfo {author} {\bibfnamefont {P.}~\bibnamefont {Dub\'e}},\ }\href
  {\doibase 10.1103/PhysRevA.96.022704} {\bibfield  {journal} {\bibinfo
  {journal} {Phys. Rev. A}\ }\textbf {\bibinfo {volume} {96}},\ \bibinfo
  {pages} {022704} (\bibinfo {year} {2017})}\BibitemShut {NoStop}%
\bibitem [{\citenamefont {Vutha}\ \emph {et~al.}(2018)\citenamefont {Vutha},
  \citenamefont {Kirchner},\ and\ \citenamefont {Dub{\'e}}}]{Vutha2018}%
  \BibitemOpen
  \bibfield  {author} {\bibinfo {author} {\bibfnamefont {A.}~\bibnamefont
  {Vutha}}, \bibinfo {author} {\bibfnamefont {T.}~\bibnamefont {Kirchner}}, \
  and\ \bibinfo {author} {\bibfnamefont {P.}~\bibnamefont {Dub{\'e}}},\
  }\href@noop {} {\bibfield  {journal} {\bibinfo  {journal} {arXiv preprint
  arXiv:1812.00973}\ } (\bibinfo {year} {2018})}\BibitemShut {NoStop}%
\bibitem [{\citenamefont {Hankin}\ \emph {et~al.}(2019)\citenamefont {Hankin},
  \citenamefont {Clements}, \citenamefont {Huang}, \citenamefont {Brewer},
  \citenamefont {Chen}, \citenamefont {Chou}, \citenamefont {Hume},\ and\
  \citenamefont {Leibrandt}}]{Hankin2019}%
  \BibitemOpen
  \bibfield  {author} {\bibinfo {author} {\bibfnamefont {A.}~\bibnamefont
  {Hankin}}, \bibinfo {author} {\bibfnamefont {E.}~\bibnamefont {Clements}},
  \bibinfo {author} {\bibfnamefont {Y.}~\bibnamefont {Huang}}, \bibinfo
  {author} {\bibfnamefont {S.}~\bibnamefont {Brewer}}, \bibinfo {author}
  {\bibfnamefont {J.-S.}\ \bibnamefont {Chen}}, \bibinfo {author}
  {\bibfnamefont {C.}~\bibnamefont {Chou}}, \bibinfo {author} {\bibfnamefont
  {D.}~\bibnamefont {Hume}}, \ and\ \bibinfo {author} {\bibfnamefont
  {D.}~\bibnamefont {Leibrandt}},\ }\href@noop {} {\bibfield  {journal}
  {\bibinfo  {journal} {arXiv preprint arXiv:1902.08701}\ } (\bibinfo {year}
  {2019})}\BibitemShut {NoStop}%
\bibitem [{\citenamefont {Davis}\ \emph {et~al.}(2019)\citenamefont {Davis},
  \citenamefont {Dub{\'e}},\ and\ \citenamefont {Vutha}}]{Davis2019}%
  \BibitemOpen
  \bibfield  {author} {\bibinfo {author} {\bibfnamefont {J.}~\bibnamefont
  {Davis}}, \bibinfo {author} {\bibfnamefont {P.}~\bibnamefont {Dub{\'e}}}, \
  and\ \bibinfo {author} {\bibfnamefont {A.~C.}\ \bibnamefont {Vutha}},\
  }\href@noop {} {\bibfield  {journal} {\bibinfo  {journal} {arXiv preprint
  arXiv:1901.06443}\ } (\bibinfo {year} {2019})}\BibitemShut {NoStop}%
\end{thebibliography}%

\end{document}